\documentclass[STIX1COL]{WileyNJD-v2}
\pdfoutput=1
\usepackage{moreverb}
\usepackage{gensymb}
\usepackage{todonotes}
\usepackage{color}
\usepackage{xcolor}
\usepackage{subfigure}
\usepackage{natbib}
\usepackage{caption}
\usepackage{float}

\usepackage[finalnew]{trackchanges} 

\newcommand\BibTeX{{\rmfamily B\kern-.05em \textsc{i\kern-.025em b}\kern-.08em
T\kern-.1667em\lower.7ex\hbox{E}\kern-.125emX}}

\articletype{Survey}%

\received{Jan-19-2021}
\revised{Jun-02-2021}
\accepted{XX-XX-XXX}

\begin{document}

\title{Critical Risk Indicators (CRIs) for the electric power grid: A survey and discussion of interconnected effects}

\author[1]{Che-Castaldo, Judy P.*}
\author[2]{Cousin, Rémi}
\author[3]{Daryanto, Stefani}
\author[4]{Deng, Grace}
\author[1]{Feng, Mei-Ling E.}
\author[5]{Gupta, Rajesh K.}
\author[5]{Hong, Dezhi}
\author[6]{McGranaghan, Ryan M.}
\author[7]{Owolabi, Olukunle O.}
\author[8]{Qu, Tianyi}
\author[3]{Ren, Wei}
\author[4]{Schafer, Toryn L. J.}
\author[9,10]{Sharma, Ashutosh}
\author[9]{Shen, Chaopeng}
\author[8]{Sherman, Mila Getmansky}
\author[7]{Sunter, Deborah A.}
\author[3]{Tao, Bo}
\author["11"]{Wang, Lan}
\author[4]{Matteson, David S.}

\authormark{Che-Castaldo \textsc{et al}}

\address[1]{\orgdiv{Conservation \& Science Department}, \orgname{Lincoln Park Zoo}, \orgaddress{\state{Illinois}, \country{USA}}}

\address[2]{\orgdiv{International Research Institute for Climate and Society}, \orgname{Earth Institute / Columbia University}, \orgaddress{\state{New York}, \country{USA}}}

\address[3]{\orgdiv{Department of Plant and Soil Sciences}, \orgname{College of Agriculture, Food and Environment / University of Kentucky}, \orgaddress{\state{Kentucky}, \country{USA}}}

\address[4]{\orgdiv{Department of Statistics and Data Science}, \orgname{Cornell University}, \orgaddress{\state{New York}, \country{USA}}}

\address[5]{\orgdiv{Halicioglu Data Science Institute and Department of Computer Science \& Engineering}, \orgname{University of California, San Diego}, \orgaddress{\state{California}, \country{USA}}}

\address[6]{\orgname{Atmosphere Space Technology Research Associates}, \orgaddress{\state{Colorado}, \country{USA}}}

\address[7]{\orgdiv{Department of Mechanical Engineering}, \orgname{Tufts University}, \orgaddress{\state{Massachusetts}, \country{USA}}}

\address[8]{\orgdiv{Department of Finance}, \orgname{Isenberg School of Management, UMASS Amherst}, \orgaddress{\state{MA}, \country{USA}}}

\address[9]{\orgdiv{Civil and Environmental Engineering}, \orgname{Pennsylvania State University}, \orgaddress{\state{Pennsylvania}, \country{USA}}}

\address[10]{\orgdiv{Department of Hydrology}, \orgname{Indian Institute of Technology Roorkee}, \orgaddress{\state{Roorkee}, \country{India}}}

\address["11"]{\orgdiv{Department of Management Science}, \orgname{Miami Herbert Business School, University of Miami}, \orgaddress{\state{FL}, \country{USA}}}

\corres{*J.P. Che-Castaldo, 2001 N. Clark St. Chicago, IL 60614. \email{jchecastaldo@lpzoo.org}}



\abstract[Abstract] 
{The electric power grid is a critical societal resource connecting multiple infrastructural domains such as agriculture, transportation, and manufacturing. The electrical grid as an infrastructure is shaped by human activity and public policy in terms of demand and supply requirements. Further, the grid is subject to changes and stresses due to diverse factors including solar weather, climate, hydrology, and ecology. The emerging interconnected and complex network dependencies make such interactions increasingly dynamic, posing novel risks, and presenting new challenges to manage the coupled human-natural system. This paper provides a survey of models and methods that seek to explore the significant interconnected impact of the electric power grid and interdependent domains. We also provide relevant critical risk indicators (CRIs) across diverse domains that may \change{influence}{be used to assess} risks to electric grid reliability, including climate, ecology, hydrology, finance, space weather, and agriculture. We discuss the convergence of indicators from individual domains to explore possible systemic risk, i.e., holistic risk arising from cross-domains interconnections. Further, we propose a compositional approach to risk assessment that incorporates diverse domain expertise and information, data science, and computer science to identify domain-specific CRIs and their union in systemic risk indicators. Our study provides an important first step towards data-driven analysis and predictive modeling of risks in interconnected human-natural systems.}


\jnlcitation{\cname{%
\author{author one},
\author{author two},
\author{author three},
\cyear{2020}, 
\ctitle{title here}, 
\cjournal{journal here}, \cvol{year;number}.}}

\maketitle

\footnotetext{\textbf{Abbreviations:} CRI}

\section{Introduction}\label{introduction section}
The electrical power grid is an example of an emerging class of human-natural systems that involve complex interdependent processes to carry out their primary functions. With the emergence of pervasive connectivity through computer/communication networks and cloud computing, these interactions are increasingly dynamic, representing myriad environmental changes and human activities. For instance, abnormal climate leads to changes in hydrology which can influence the inputs to hydroelectric production \citep{uria2021, voisin2016vulnerability, Scott_2013}. Similarly, space weather is a well-known source of disturbance to the power grid \citep{Boteler_2001}, while wildlife can impact its reliability \citep{doostan_statistical_2019, polat_overview_2016,NRECA_2016, maliszewski_environmental_2012, chow_analysis_1995}. Agricultural production \citep{lewis2020short, hicks2014energy}, water access, distribution and groundwater pumping \citep{pump_hydro} serve on the demand side of the grid \citep{gonzalezmultipurpose_res}. In turn, the thermal power plants are large water users and, in unfortunate events, could pose threats of various forms to ecosystems \citep{marques_wind_2019, Falke2011}.

Many of these subsystems are shifting due to the stresses of climate change as well as integration of renewable energy sources and electric vehicles causing large variability in electricity use and availability. The hydrologic cycle is undergoing intensification with more frequent floods and droughts. Droughts, in particular, increase the propensity for wildfires which directly negatively impact the power grid through power supply interruptions and physical destruction of the grid \citep{Allen-Dumas2019, dian2019integrating}. On the other hand, droughts can decrease hydropower \citep{Gleick2015} and are known to lead to fluctuations in water and food commodity prices \citep{badiani2018electricity}. Given the increasing variability of and stress on these subsystems, there is an ever-increasing need to understand risks from a holistic point of view. We can use quantifiable risk measures to design strategies to improve resilience of the electric grid during periods of vulnerability. 

Several concepts seek to characterize a system's functional response during periods of vulnerability, describing how the system minimizes losses, how it maintains desired functions, and its rate of recovery \citep{galaitsi_need_nodate}. Vulnerability can be defined as the problems a system faces to maintain its function after an accumulation of risk leads to an unwanted event \citep{sperstad_comprehensive_2019}. These vulnerabilities can be a result of the interdependencies between system components, each with its own sets of risk that can compound into systemic risks \citep{hynes_bouncing_2020,golan_trends_2020}. Systemic risk involves a system, e.g., a collection of interconnected domains, through which losses, insolvency, and natural disasters can quickly propagate resulting in systemic distress \citep{Billio2012}. These interdependencies between diverse components in a system can provide profound insights into the health and risk states of the system as a whole, yet there is a dearth of rigorous definition, understanding, and meaningful review of existing indicators of risk for society's most important systems \citep{galaitsi_need_nodate, golan_trends_2020,brand_focusing_2007,raoufi_power_2020,izadi_critical_2021}. Additionally, despite a recent surge in studies of intra-system resiliency, there is still a gap in studies of inter-system networks across interdependent sectors \citep{golan_trends_2020, buldyrev_catastrophic_2010}.

In this paper, we fill the gap with a survey of risk indicators for the electric power grid system, identifying those indicators across a range of domains that must be considered to improve the resiliency of the power grid. Definitions of `resiliency' often overlap with other terms that characterise systemic functional responses \citep{raoufi_power_2020, galaitsi_need_nodate, brand_focusing_2007, maliszewski_environmental_2012, akhgarzarandy_optimal_2021, izadi_critical_2021}. Resiliency encompasses multiple components that describe the system's reaction, response to, and recovery from disturbance \citep{raoufi_power_2020, voropai_electric_2020, noauthor_definition_2018}. In this survey, we concentrate on the reactive components of power system resilience \citep[following][]{raoufi_power_2020} by focusing on the electric grid system's ability to withstand low-frequency, high-impact disasters efficiently while minimizing interruption in electricity supply. Resiliency emphasizes the addition of risk from acute events and builds upon system reliability which focuses on inherent risk from recurring events \citep{akhgarzarandy_optimal_2021, izadi_critical_2021, noauthor_definition_2018,ciapessoni_defining_2019, jufri_state---art_2019,sperstad_comprehensive_2019}.

Previous literature has demonstrated the importance of studying interdependencies of critical infrastructure, which are the frameworks underlying complex, adaptive systems that provide institutional services essential for the economy, government, and society as a whole \citep{rinaldi_identifying_2002}. Although managers of critical infrastructure have taken approaches to prepare for frequent and predictable disturbances, termed as chronic threats \citep{galaitsi_need_nodate}, they still lack the capacity to account for and recover from extreme events with low probabilities of occurrence \citep{kurth_lack_2020}.
\add{Following one definition of risk from the Society of Risk Analysis} \citep{aven2018society}\add{, we define risk as the potential for realization of unwanted, negative consequences of an event. We take a rigorous approach to risk measurement, defining the term Critical Risk Indicator (CRI) as quantifiable information specifically associated with an unfavorable state, which may be a \textit{disastrous} activity (cumulative) or a \textit{catastrophic} event}  \citep[acute;][]{galaitsi_need_nodate} \add{that is devastating and leads to ruinous losses}. We take a power supply interruption in the power grid as the realization of risk in the context for CRI development in this study. Our subsequent CRIs \add{can be used to} quantify risks that can overwhelm the coping capacity of the energy system \citep{sperstad_comprehensive_2019, abedi_review_2018,zio_challenges_2016} leading to undesirable emergent behaviors \citep{rinaldi_identifying_2002} and critical, wide-spread, multi-domain vulnerability. With a more complete understanding of the full inter-system network, grid managers can better prepare for, anticipate, and detect a more holistic range of potential risks to the electrical grid and as a result, sustain critical operations and speed up its recovery in the face of future extreme events \citep{ciapessoni_defining_2019}.

This paper goes beyond surveying traditional risk indicators used in the electric energy sector and captures the risk indicators across diverse domains. We broaden the interdependencies of human-centred critical infrastructure to the human-natural interface, highlighting the linkages between geophysical and ecological processes, economics, and community services. Specifically, we survey and provide methodologies for critical risk indicators (CRIs) in electric energy, finance, climate, ecology, space weather, hydrology, and agriculture domains. In Section \ref{Critical Risk Indicators (CRIs) by Domain}, we provide an overview of the human-natural domains, discuss how they impact electric power grid resiliency, and present datasets available for developing CRIs in each domain. Section \ref{energy section} describes the power grid and some existing electric energy CRIs. We conclude in Section \ref{systemic risk} with a discussion about converging CRIs from individual domains to explore systemic risk. We describe how the human-natural domains are also interdependent, making it necessary to consider systemic risk, or the holistic risk arising from cross-domain interconnections. We borrow the concept of systemic risk and systemic risk measures from finance literature. In finance, systemic risk is the risk that the collapse of one financial institution can lead to the cascade of failures in other financial institutions, and the financial system as a whole \citep{Billio2012}. In our case, measuring systemic risk would capture the health of the interconnected human-natural system and the interdependencies between CRIs in each of the human-natural domains; we refer to the trans-domain systemic risk measures as Systemic Risk Indicators (SRIs). Finally, we propose a compositional approach to develop network-based SRIs by dynamically modeling domain-specific CRIs and quantifying inter-domain connectivity and risk spillovers. Future work will focus on expansion and full implementation of our network approach to measure systemic risk and identify mitigation targets to improve electric power grid resiliency. Although we focus on the electric power grid in the United States, the approach we present here for identifying domain-specific CRIs and their union in SRIs lends itself to broader application for the exploration of the complex interconnectedness and systemic risks in human-natural systems.

\section{Critical Risk Indicators (CRI\lowercase{s}) by Domain} \label{Critical Risk Indicators (CRIs) by Domain}

\begin{figure}[h]
    \centering
    \vspace{-10mm}
    \includegraphics[width = 0.5\textwidth]{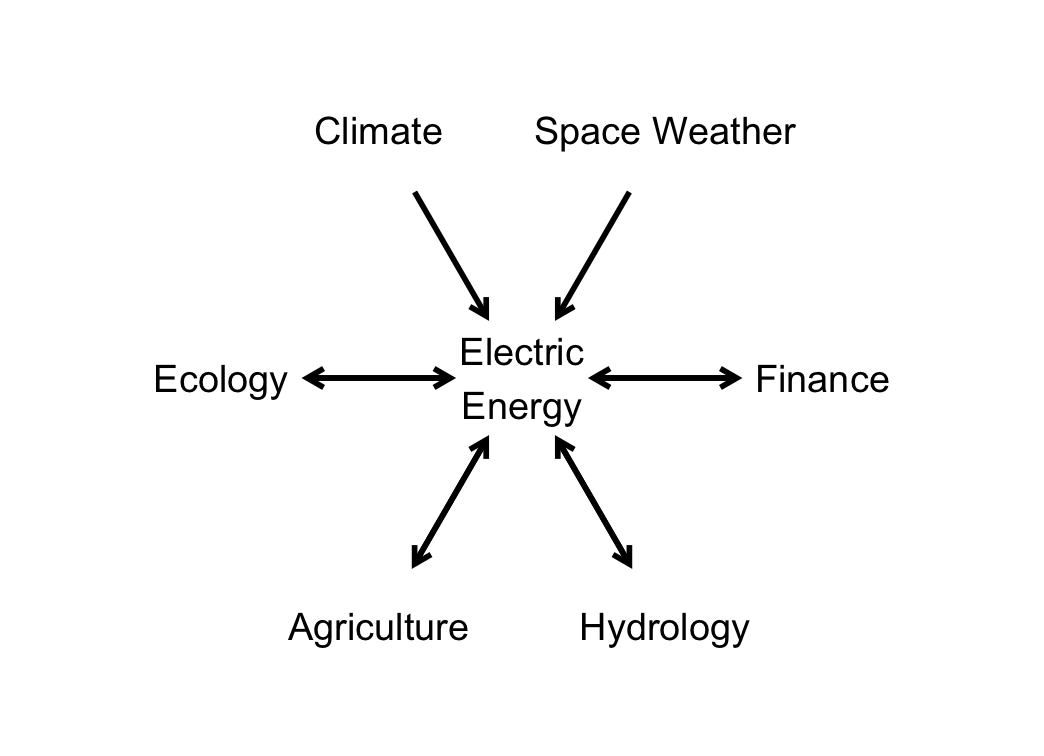}
    \vspace{-7mm}
    \caption{Nexus of interconnections among different human-natural domains and the electric energy domain.}
    \label{fig:cri_dag}
     \vspace{-10mm}
\end{figure}

Disruption to the electric power grid is a systemic event which is a result of interdependencies between the following human-natural domains: climate, hydrology, agriculture, ecology, space weather, and finance. Vulnerability in each of the six domains spills over to the electric energy domain, and in some instances the relationship is reversed with electric energy domain vulnerabilities cascading to other human-natural systems. Given that a Critical Risk Indicator (CRI) is an entity that relates to a specific catastrophic outcome, each section will provide an overview of existing CRIs within that particular domain in the context of \textit{disruption to the electric power grid.}  We describe each domain, provide a background on the connection between each domain and risk to the electric power grids, survey top CRIs for each domain that relate to electric power grid, and provide discussion of interconnections with other domains. Table \ref{summary cris table} summarizes the CRIs. Figure \ref{fig:cri_dag}provides a nexus of interconnections among different human-natural domains and the electric energy domain.

\subsection{Climate} \label{climate section}

\textit{Connection between climate and the electric power grid}: 
Climate change and extreme weather (e.g., droughts, \citealp{voisin2016vulnerability}, and strong winds, \citealp{wanik2017using}) are a leading cause of power supply interruptions. Additionally, climate-related events such as wildfires \citep{dian2019integrating} and fallen vegetation can cause up to 90\% of storm-related power interruptions
\citep{wanik2017using}. Drought events \change{create}{disrupt} not only \change{an}{the} electricity supply \remove{risk} (e.g., reduced capacity of water-cooled thermoelectric plants \citep{voisin2016vulnerability}) but also \remove{a} storage capacity\remove{risk leading to an electricity availability risk}, as pumped hydropower accounts for 95\% of all utility-scale energy storage in the US \citep{pump_hydro}. Climate change and extreme weather \change{is}{are} also associated with \add{unstable} electricity supply \remove{risk }from reduced generation from variable renewable energy, such as solar \citep{feron2021climate} and wind \citep{lin2012assessment}.
Heat waves and cold spells are likely to increase power demand to cool and heat buildings. Below freezing temperatures, resulting in ice accumulations, \change{could pose a risk to}{have the potential to damage} electric grid infrastructure \citep{Allen-Dumas2019}. Persistent high temperatures led to increased electricity consumption and increased power supply interruptions \citep{CaliforniaISO2021}. 

\textit{Existing CRIs for climate-power grid connections}: 
The main objective in climate science is to characterize, understand and consequently try to predict anomalous climate events. To define what is abnormal, one must first define "normal" conditions. In climate science, these normal conditions are defined as the seasonal cycle, typically defined, given monthly data, as the 30-year average of each month of the year. Then anomalies are simply the departure for a given month and year from that corresponding month average, and thus climate science focuses on yearly time frequencies or lower. \change{Now t}{T}his broad concept can take other forms: for instance finer time resolution than the month (e.g., daily, 5-daily, 10-daily); the data can be aggregated at larger time resolution than its step (e.g., monthly 3-month long seasons averages, a running average in other words); or anomalies can be standardized or normalized according to different techniques.

\begin{figure}
\caption{Number of cold days (0\degree C or less) expressed in anomalies with respect to the 1981-2010 average, from NASA MERRA2 Reanalysis. For January 2019 (map) and for 88.125\degree W, 37\degree N (bar plot)} 
\centering
\subfigure{\includegraphics[width=80mm]{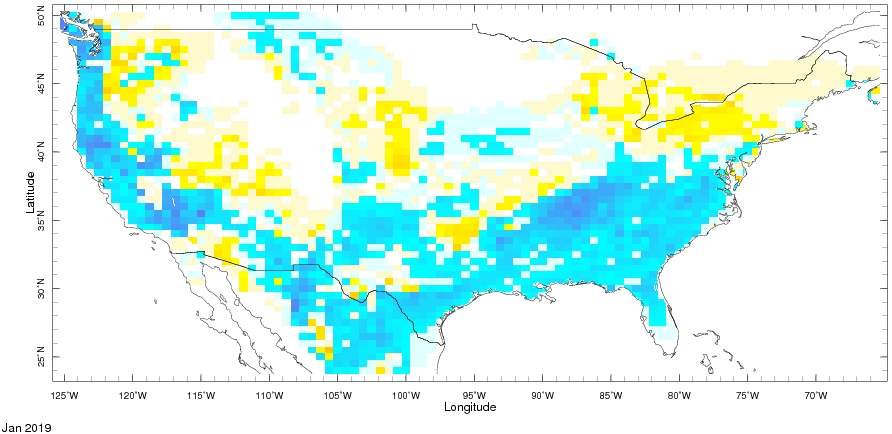}}
\subfigure{\includegraphics[width=80mm]{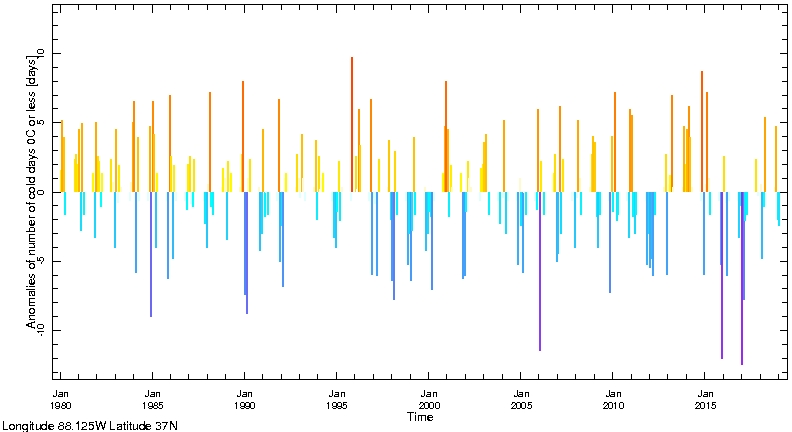}}
\subfigure{\includegraphics[width=150mm]{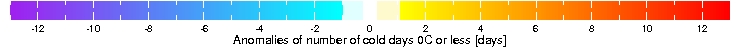}}
\label{climFig}
\end{figure}

In light of this, we consider the following climate domain CRIs that directly relate to the electric power grid: 

\begin{enumerate}

    \item \textit{Monthly temperature/precipitation anomalies} consist for a given month and year (at any given spatial entity) of the difference between the temperature/precipitation of that month/year with the average over at least 30 years of the temperature/precipitation for that month. High anomalies (negative or positive) would be \change{a}{an} \remove{risk }indicator \change{for}{of} stress towards the power grid. Long term year-to-year relationship can be assessed between temperature/precipitation anomalies and power supply interruptions, in particular when looking at the same time period of the year (e.g., hot summers are known to cause more power supply interruptions than cool ones \citep{CaliforniaISO2021}). But also sequences of adverse conditions from a season to another could be assessed (e.g., cold winters followed by hot summers).

    \item \textit{Standard Precipitation Index (SPI)} is an index used to characterize drought on a range of timescales.  It characterizes drought or abnormal wetness at different time scales \citep{GuyMerlinGuenang2014}. SPI is also related to propensity of wildfires that directly affects power supply interruptions and electric grid infrastructure.
    
    \item \textit{Anomalies of number of days a criterion is met} (e.g., $>1$mm; $\leq 0${\degree}C). For example, cooling degree days are summations of positive differences between the daily temperature and a reference base temperature during a season of interest \citep{U.S.EnergyInformationAdministration2020}. For instance summing up, through days, temperature above 20{\degree}C during the summer, as an indicator of how much cooling power is necessary to maintain desired temperature in buildings. In another example, one could rely on daily data to build monthly anomalies of number of days below a critical temperature (e.g., freezing point, i.e. 0{\degree}C -- see Figure \ref{climFig}) in a month. Such a CRI could be more tailored to relate to power supply interruptions in the winter.
    
\end{enumerate}

\textit{Available datasets for calculating CRIs:} There exists a number of datasets of daily or monthly precipitation or temperature, over several decades, that cover the United States, that allow to calculate the CRIs described above. A few examples that we have used or planned on using follow. Climatology Lab's gridMET \citep{https://doi.org/10.1002/joc.3413} dataset has daily precipitation and temperature data (and more) from 1979 to now and at 1/24th degree of spatial resolution, over the United-States. GPCP V2.3 Monthly Analysis Product \citep{atmos9040138} has monthly precipitation from 1979 to now at 2.5 degree spatial resolution, over the globe. The Modern-Era Retrospective Analysis for Research and Applications, Version 2 (MERRA-2) \citep{TheModernEraRetrospectiveAnalysisforResearchandApplicationsVersion2MERRA2} has daily precipitation and temperature date from 1980 to now at about a half degree spatial resolution over the globe. The Climate Prediction Center's (CPC) Unified Gauge-Based Analysis of Daily Precipitation over the CONUS \citep{AGaugeBasedAnalysisofDailyPrecipitationoverEastAsia, https://doi.org/10.1029/2007JD009132,Chen_etal2008} has daily precipitation data from 1948 to now at a quarter degree spatial resolution over the CONUS.

\subsection{Hydrology} \label{hydrology section}
\textit{Connection between hydrology and electric power grid:} Hydropower plants generate about 6.7\% of total electricity generation in the United States (US) and account for about 38\% of electricity generation from renewable energy \citep{uria2021}. Globally, the percentage of electricity from hydropower is 16\% with this fraction above 90\% in some countries, e.g., Albania, Paraguay, Nepal, Congo, Ethiopia and Norway \citep{WB2021}. These countries are much more heavily influenced by the natural hydrologic system and annual precipitation. The hydropower production relies on the water available to flow through the turbines that generate electricity. The reservoirs of hydroelectric dams store water that is released through the turbine to produce electricity to meet baseload as well as peak load demands. Thus hydrology is directly linked to hydropower production through the amount of water flowing into the reservoir and its fluctuations under extreme hydrologic events (e.g., droughts and floods). The outflow from a reservoir is controlled by the reservoir release policies which may be influenced by electricity prices \citep{GAUDARD2014172,KANAMURA20071010}. In many other markets, however, the release policies need to consider the \change{risk of}{variability in} supply, along with other factors like ecosystem flow, and could have a large impact on electricity prices \citep{Doorman1583739, WOLFGANG20091642}. Drought is a hydrologic phenomenon that starts with a period of less precipitation compared to historical normal (meteorological drought), and if precipitation deficit sustains over an extended period, it results in reduced soil moisture (agricultural drought) and surface water (i.e., lakes, reservoirs, rivers, and wetlands) deficit (hydrological drought). Prolonged drought events affect water storage in these reservoirs, and hence limit the ability to generate electricity. The past drought events had substantially impacted the regional/national hydropower productions in different countries. For example, the 2011-2015 California drought resulted in below-average hydropower production that added an economic cost of \$2.0 billion \citep{Gleick2015}. Further, the fossil fuel-based electricity generation was enhanced to meet the electricity demands in California, leading to a 10\% increase in CO$_2$ emission from power plants \citep{CARB2015}. In Brazil, the 2012-2015 droughts were already causing lowered hydropower productions and elevated thermal dispatches \citep{Zambon2016}.

\textit{Existing CRIs for hydrology-power grid connections:}
The hydrologic risk (i.e., \add{potential for} hydrological drought) is quantified based on prolonged abnormally low streamflow and groundwater depletion. CRIs in hydrology are (i) the drought indices that quantify the deviation in water availability (surface water or groundwater) compared to long-term historical normal; (ii) multi-month streamflow outlook. \add{Existing CRIs in hydrology include}:

\begin{enumerate}
    \item \textit{Streamflow:} Streamflow, when put in historical context, is a useful indicator for hydrologic risks (i.e., \add{potential for} drought or flood). The values of streamflow are converted in percentiles and are compared to historical observations during the same period of the year based on a threshold (e.g., 10th \%-ile of past decades distribution). Apart from present streamflow conditions, to provide a useful tool to forecast risk, multi-month streamflow outlook can be estimated based on machine models, of various mechanisms and climate outlooks, e.g., see some preliminary work in \cite{Feng2020,Ouyang2021} which can be extended to multi-month outlook.
    \item \textit{Drought indices:} Several drought indices have been developed over the years to identify droughts and to quantify the drought intensity/severity \citep{Svoboda2016}. Palmer Drought Severity Index (PDSI) and Standardized Precipitation Index (SPI) are the most widely used drought indices. SPI is recommended by the World Meteorological Organization (WMP) and requires only monthly precipitation data. SPI is a meteorological drought index, but it can be computed for multiple timescales (e.g., 3, 6, 12, 24 months) that enables us to examine other types of droughts (agricultural or hydrological). PDSI uses readily available temperature and precipitation data to estimate relative dryness. It is a standardized index that generally spans -10 (dry) to +10 (wet).
    \item \textit{Groundwater levels:} Groundwater depletion rates provide information on excessive pumping activities for irrigation during the drought. The observations of groundwater wells can be used as a CRI that accounts for the change in groundwater table depth or the fraction of dry wells. The water stored in a region can also be reflected from satellite-based observations of terrestrial water storage \citep{Li2012,Sun2012}, but the downside of this kind of observations is their very coarse spatio-temporal resolutions. On the other hand, access to groundwater requires energy \citep{CHEN20191033,Siddiqi13} and could, in turn, affect the grid.
\end{enumerate}

\begin{figure}[htp]
\centering
\caption{Time series of normalised streamflow at different gages in California. The red region highlights streamflow reduction during the 2011-2016 drought.}
\includegraphics[width=.8\textwidth]{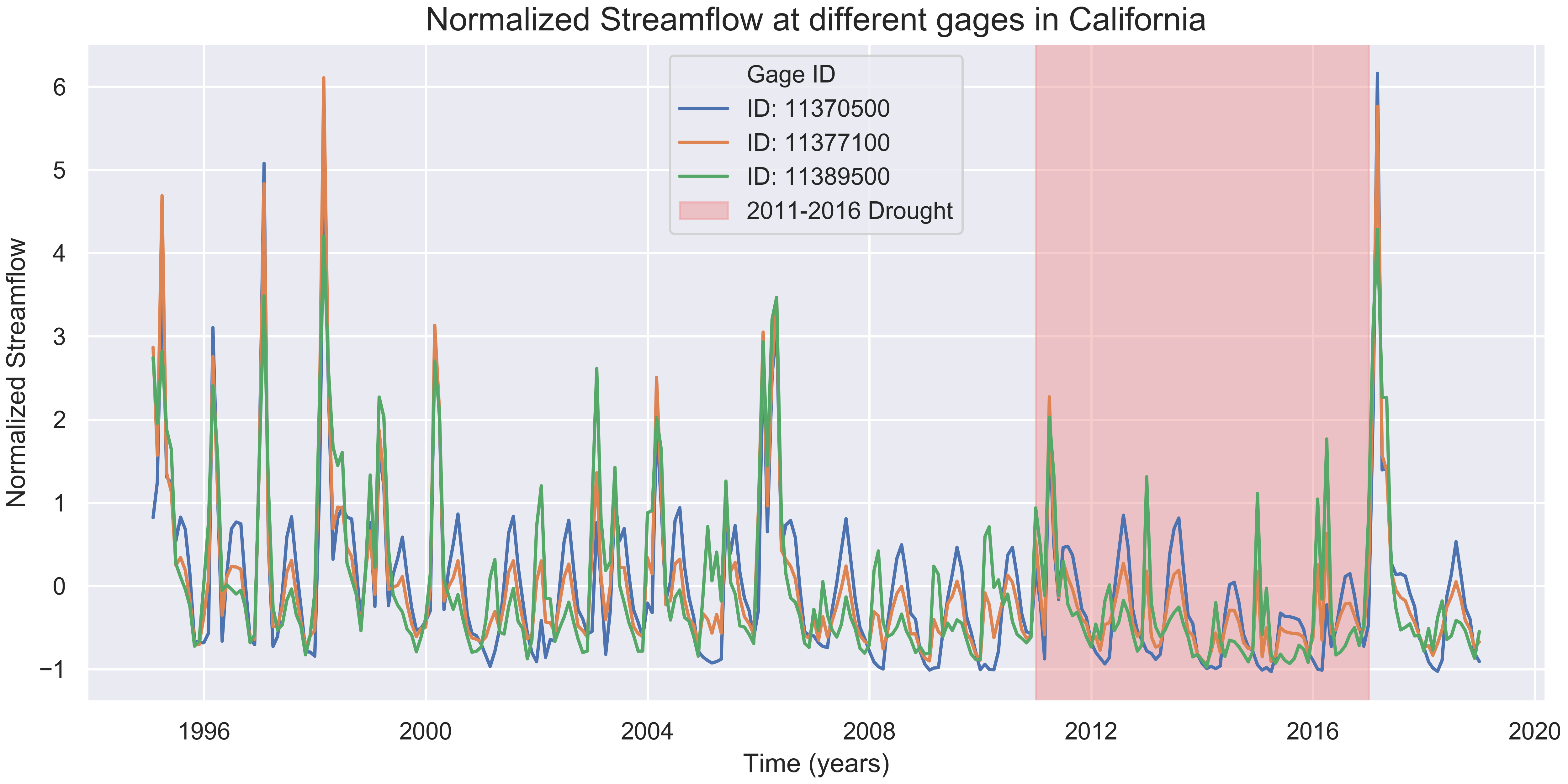}
\label{Normalized Streamflow}
\end{figure}

\textit{Available datasets for calculating CRIs:} Daily streamflow observations are available for all major rivers in the US from the United States Geological Survey (USGS) National Water Information System (NWIS)\footnote{https://waterdata.usgs.gov/nwis/rt}. GAGES II (Geospatial Attributes of Gages for Evaluating Streamflow, version II) dataset provides a large set of geospatial data for 9322 gage sites across the US including environmental features (e.g., climate – including historical precipitation, geology, soils, topography) and anthropogenic influences (e.g., land use, road density, presence of dams, canals, or power plants). Figure \ref{Normalized Streamflow}shows the normalized streamflow for some gages in California over the period 1995-2019 and highlights the reduction in streamflow during the 2011-2016 drought. 

The USGS NWIS provides data on groundwater well observations for sites across US\footnote{https://waterdata.usgs.gov/nwis/gw}. Additionally, different states have networks of a large number of monitoring wells. For example, the Department of Water Resources, California provides groundwater data for thousands of wells in the state on the Water Data Library (WDL)\footnote{http://wdl.water.ca.gov/}. 

\subsection{Agriculture} \label{agriculture section}
\textit{Connection between agriculture and power grid:}  
Many agricultural activities (e.g., pumping for irrigation, supplying water for livestock) benefit from the availability of electricity. Electrification in the rural regions where most agricultural activities occur leads to increases in agricultural production \citep{lewis2020short}. Rural electrification is also associated with increased irrigation use in the western region of the US and substantial increases in the average farm size \citep{lewis2014value}. Those increases also correspond with advances in power transmission technology, which reduces the constraints on where power plants can be located \citep {lewis2014value}. 
Electricity allows for expansion in agricultural activities through two mechanisms. First, electricity allows mechanization of equipment such as grain mills and electric dryers \citep{shrestha2005application}. 
Second, electricity allows extended working hours, which again leads to higher production capacity. 

Yet as farm productivity becomes more dependent on grid electricity, it  also means that there will be losses if electricity is not available. Security breaches in electric power transmission systems (e.g., outages, transport, etc.) resulted in several blackouts in the US during the late 2000s \citep{arianos2009power}, which incurred large losses in agriculture, particularly when the power disruptions occurred during the periods of peak electricity demand \citep[August-September harvest,][]{lewis2020short}. A four-hour duration of electricity interruption cost (USD 1.94 kW-1) is relatively higher in the agricultural sector compared to coal (USD 0.07 kW-1) or metal mining (USD 0.11 kW-1) based on 1994 currency value \citep{badiani2018electricity}. If the food industry, as an extension of agriculture, is included, the interruption cost jumped to USD 50.52 kW-1 due to spoilage \citep {balducci2002electric}.

Agriculture acts as both a supplier and a consumer of electricity. \change{When working as a supplier}{As an energy source}, the amount of residue generated from agriculture \change{becomes the CRI for}{influences} electricity \add{supply and} generation. In 2016, biomass and waste fuels supplied approximately 2\% of total electricity generation in the US \citep[71.4 billion kWh,][]{Mayes2017}. Wood solids, which come from sources like logging and mill residues, accounted for nearly 33\% the electricity generated from biomass and waste \citep{Mayes2017}. To generate electricity, they can be burned directly in steam-electric power plants or be converted to a gas. The gas then can be burned in steam generators, gas turbines, or internal combustion engine generators \citep{USDOE2020}. 

On the other hand, when acting as a consumer of electricity, crop production and cropland area, especially those requiring irrigation, are highly dependent on the steady supply of energy.  In 2012,  US crop production obtained about 20\% of its energy requirement from electricity \citep{hicks2014energy}. The agriculture-heavy regions of Nebraska (i.e., rural south and west) have one of the highest average electricity prices in the state \citep{Brown2014short}. Demand for irrigation can be costly, because of two main reasons. First, it is expensive to connect dispersed farmlands to the electric grid and second, it is also expensive to provide enough capacity available to meet seasonal irrigation load \citep{Brown2014short}.

\begin{figure}
    \centering
    \caption{MODIS EVI values across the US croplands in normal year 2010 (left) and drought year 2012 (right). The two EVI maps were calculated from the MOD09A1 Version 6 product with a 500m spatial resolution and an 8-day temporal resolution.}
    \vspace{-6mm}
    \includegraphics[width=15cm]{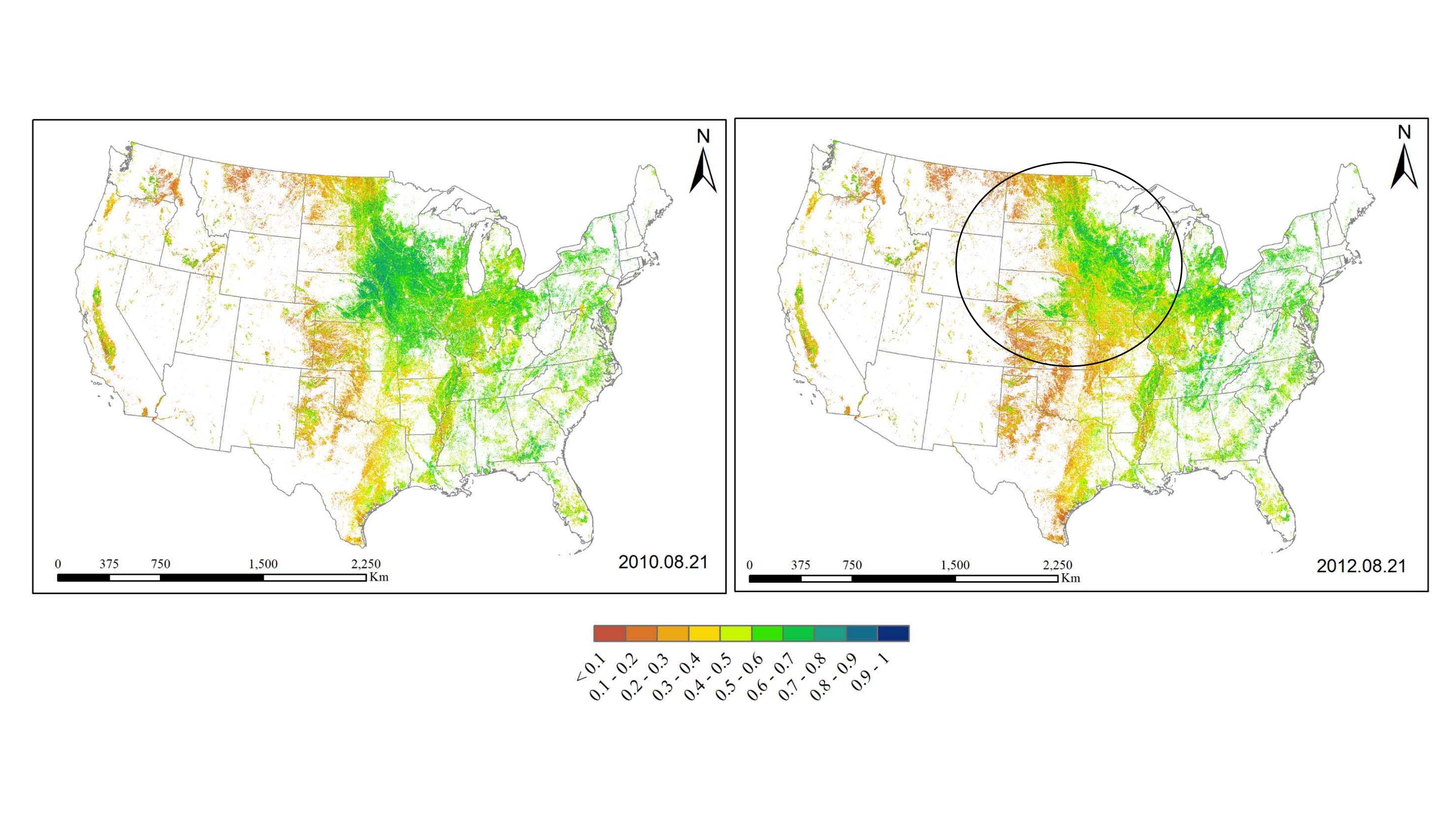}
    \vspace{-8mm}
    \label{agricultureFig}
\end{figure}

\textit{Existing CRIs for agriculture-power grid connections:} 
\add{Agriculture risk is related to catastrophic declines in crop biomass production and vegetation index, as well as a possibility of not meeting irrigation demand. The agricultural sector is highly reliant on the electric energy sector, and thus power supply interruptions can exacerbate risk in the agriculture sector.  Relevant CRIs for agriculture-power grid connections include}:  

\begin{enumerate}
    \item \textit{Irrigation demand:} The irrigation demand is a useful indicator for evaluating power grid risk caused by agriculture as a consumer of electricity.  The larger irrigation demand requires more  energy capacity support, relating to irrigation area and electricity price.
    \item \textit{Crop biomass production:} The total biomass production and its reduction are \change{useful}{agricultural} indicators \add{related} to \change{predict agriculture}{the} supply \change{risk for whole}{and generation of} electricity\remove{ generation}.
    \item \textit{Vegetation Index:} The Enhanced Vegetation Index (EVI) is an 'optimized' vegetation index designed to quantify vegetation greenness (Figure \ref{agricultureFig}). The EVI represents plant growth status and relates to irrigation demand and final biomass production. 
\end{enumerate}     

\textit{Available datasets for calculating CRIs:} Data on the land-use type and irrigated croplands are available from the United States Department of Agriculture, Economic Research Service\footnote{ https://www.ers.usda.gov/data-products/major-land-uses/; https://www.ers.usda.gov/data-products/irrigated-agriculture-in-the-united-states/}. Electricity price is available from US Energy Information Administration (note that the agricultural sector is considered as an industrial sector \citep{Brown2014short}. Electricity power generation from biomass is also available from the US Energy Information Administration\footnote{https://www.eia.gov/electricity/monthly/}.


\subsection{Ecology} \label{ecology section}
\textit{Connection between ecology and the electric power grid}: 
Biotic components of the environment can both negatively impact and be impacted by the electrical power grid. Vegetation frequently interferes with overhead power lines, particularly through tree falls, which are more likely to occur during severe weather events \citep{wanik2017using, maliszewski_environmental_2012}. Small mammals and birds cause a large proportion of disruptions to the electricity supply and damage to infrastructure \citep{chow_analysis_1995, doostan_statistical_2019}. Negative impacts on wildlife result from coexistence and attraction to electric infrastructure for use as hunting perches, nesting structures, and highways for travel \citep{NRECA_2016}. They include electrocutions and collisions with power lines \citep{polat_overview_2016}, and reduction in the quality and amount of species’ habitat taken up by electrical grid infrastructure \citep{marques_wind_2019}. 

\change{T}{While also accounting for infrastructure design and wildlife protection strategies, t}he potential for species-power grid interactions depends on both the abundance and distribution of the interacting species. For example, higher densities of individual animals would increase the likelihood for collisions, and power lines located along migration routes would pose a greater threat to birds than those away from the routes. Although spatial distribution data are available for many taxonomic groups, they typically consist of static maps of species range areas, which may not be useful for detecting associations with catastrophic events in time. In contrast, abundance data are usually time series of repeated counts over time. Moreover, species abundance is directly related to a critical risk in ecology - the risk of biodiversity loss \citep{ipbes_2019}.  

\begin{figure}
\caption{Ecology CRIs calculated using USGS Breeding Bird Survey annual abundance data for 421 species in North America from 1993-2017 (left), and monthly estimated relative abundance for the Red-bellied Woodpecker (\textit{Melanerpes carolinus}) in Massachusetts from 2005-2018 based on eBird occurrence data (right).} 
\centering
\subfigure{\includegraphics[width=75mm]{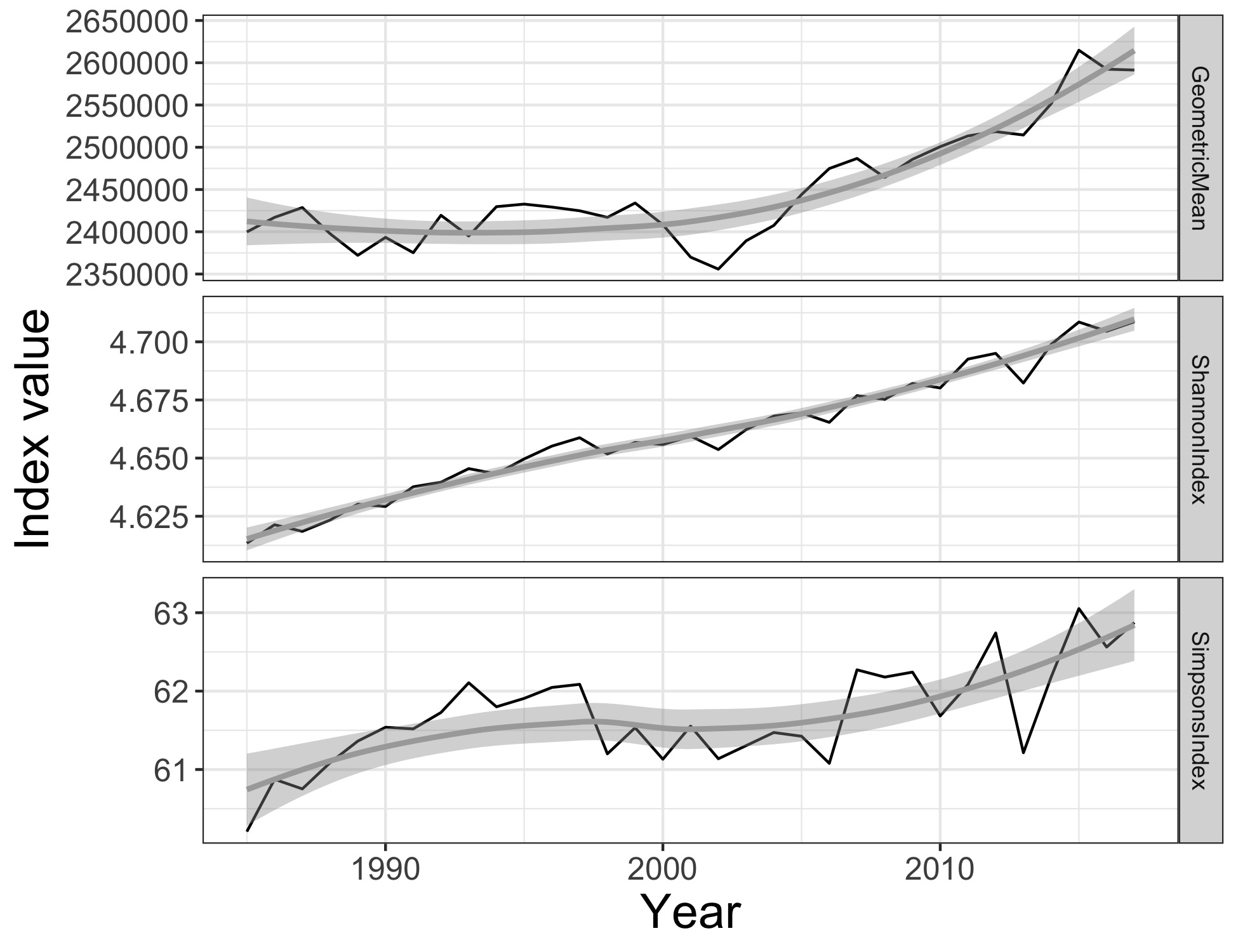}}
\subfigure{\includegraphics[width=95mm]{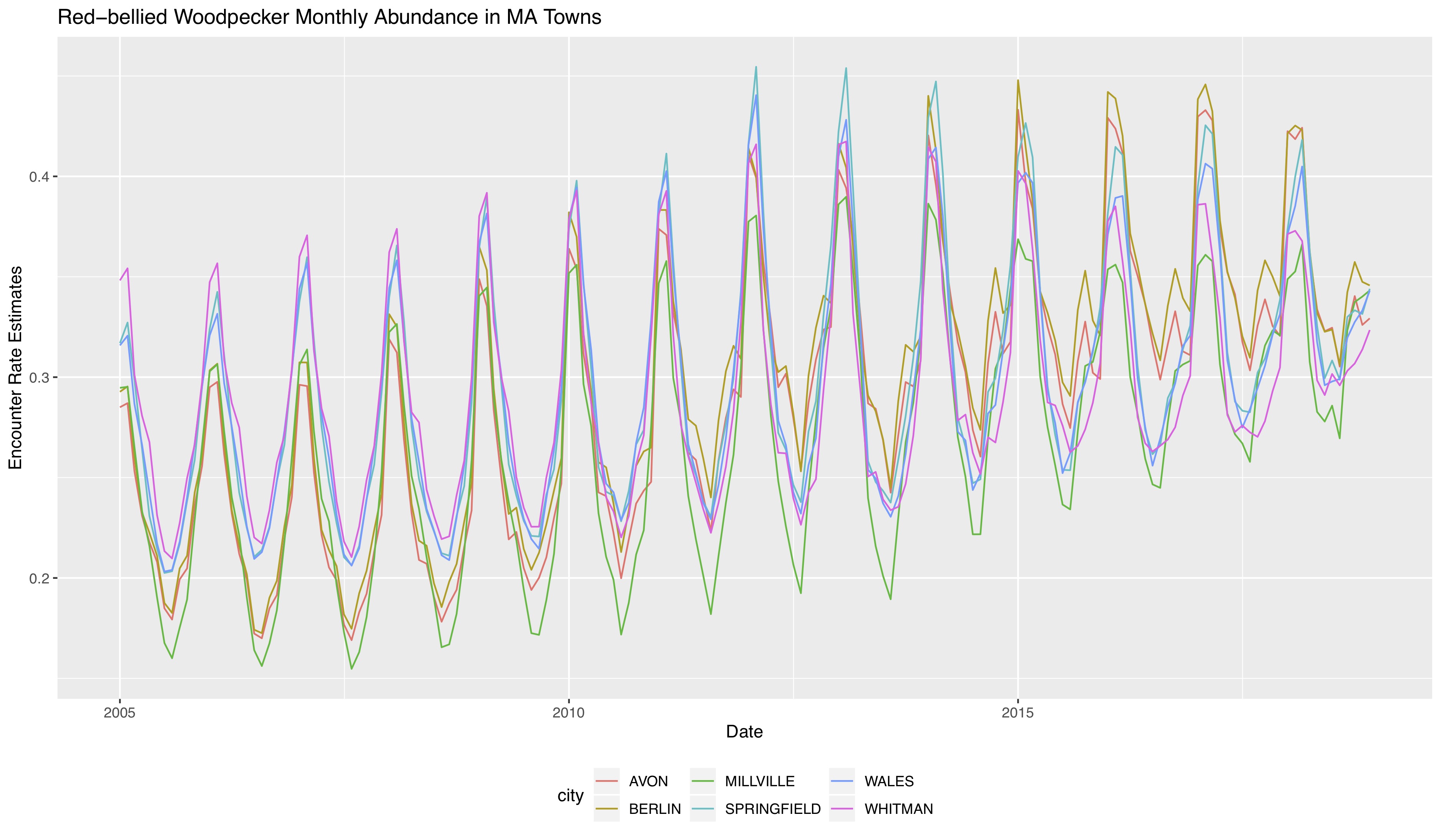}}
\label{ecologyFig}
\end{figure}

\textit{Existing CRIs for the ecology-power grid connection}: 
The critical risk of biodiversity loss can result from cumulative declines in species abundance, as well as the catastrophic event of species extinction. Potential indicators for the risk of biodiversity loss include direct measures of species abundances, as well as:

\begin{enumerate}
    \item \textit{The Living Planet Index (LPI) \citep{collen_monitoring_2009}} is one of the most comprehensive indicators of global biodiversity status. LPI is calculated as the geometric mean of population abundance trends across all species worldwide with existing abundance time series data. The geometric mean of relative abundances has empirical \citep{buckland_monitoring_2005} and theoretical \citep{mccarthy_linking_2014} support for being appropriate for assessing the risk of biodiversity loss over time.

    \item \textit{Community composition metrics} are also used to measure change in biodiversity over time \citep{buckland_monitoring_2005, morris_choosing_2014}. They include species richness (number of species) and metrics of diversity and evenness (e.g., Shannon’s or Simpson’s index; Figure \ref{ecologyFig}, left panel).
\end{enumerate}

\textit{Available datasets for calculating CRIs}: 
Due to the amount of effort and training required, the majority of abundance datasets are short-term (several years), and collected at single or few sites for single or few species. The USGS Breeding Bird Survey (BBS) is the most extensive existing dataset on animal abundance, with consistent data for a large number of bird species (\textasciitilde 400) and excellent spatial (North America, by state or by Bird Conservation Region) and temporal (annual, 1966-2017) coverage. BBS data are gathered through point count surveys along specified routes using a standardized monitoring protocol, conducted by qualified volunteers. The dataset consists of yearly, species-specific abundance indices estimated from a hierarchical trend model that accounts for differences among routes and observers \citep{sauer_first_2017}. Raw survey data are also available. A recent study demonstrated the potential of using BBS data for quantifying the magnitude of biodiversity loss \citep{rosenberg_decline_2019}.

Similarly large-scale, consistent abundance datasets do not exist for other species that may interact with the electrical grid, such as squirrels and other small mammals. It may be possible to derive proxies of relative abundance using occurrence datasets such as the Global Biodiversity Information Facility (GBIF). Occurrences differ from abundances because they are sightings or observations of a species at particular locations and times, and therefore are affected by detection probabilities and observer effort in addition to actual species abundances. However, occurrence data have finer spatial and temporal scales and may be more versatile for aligning with other domain data. For example, eBird has occurrence data collected by citizen scientists via semi-structured protocols that can be modeled to account for detection and effort and estimate relative abundance \citep{strimas-mackey_best_2020} (Figure \ref{ecologyFig}, right panel).

Finally, datasets on the abundance of vegetation that can interact with the electrical grid include the remotely-sensed normalized difference vegetation index (NDVI), which is a measure of vegetation cover with resolution of 250m and every 16 days. The Soil Adjusted Vegetation Index (SAVI) is derived from NDVI and was previously used to successfully predict power supply interruptions \citep{maliszewski_environmental_2012}. 

\subsection{Space Weather} \label{space weather section}
\textit{Connection between space weather and the electric power grid}: During periods of enhanced space weather activity, a series of physical processes beginning with the launch of a coronal mass ejection (CME) or a high speed stream (HSS) from the Sun gives rise to intense electric currents reaching millions of Amperes surrounding the Earth, which then become electric currents on the ground flowing through electrical transmission lines. This phenomenon, known as Geomagnetically Induced Currents (GICs), can disrupt the operation of high-voltage power grid transformers via overheating and generation of harmonics, potentially leading to failures.

The most fundamental quantity that connects space weather and the electric power grid is the horizontal electric field on the Earth’s surface (geoelectric field). The geoelectric field determines the magnitude of GICs that flow on power transmission networks \citep{Boteler_2001, Pirjola_2000}. GICs arise from a series of interactions, beginning with the solar cloud of plasma interacting with the Earth’s magnetic field, creating currents in space and in the upper atmospheric region known as the ionosphere, which produces the electric field on the ground through magnetic induction. However, knowledge of many aspects of this chain is limited, especially during extreme storms \citep{Ngwira_2015, Ngwira_2018}.

\textit{Existing CRIs for space weather-power grid connections}: In the space weather domain `critical risk indication' has several potential definitions, including:
\begin{enumerate}
    \item Specification of periods when the Sun is particularly active (proxies: sunspot number, location in the 11-year solar cycle.)
    \item Identification of `geomagnetically effective' periods in solar wind data \citep{SCHRIJVER2015} (important parameters: magnetic field, particularly the north-south component, velocity, density);
    \item Extent of the coupling between the solar wind and the magnetosphere by coupling function proxies: the Borovsky coupling function \citep{Borovsky_2013} and the Newell coupling function \citep{Newell_2007}; and 
    \item Activity of the current systems in the Earth's upper atmosphere proxies: the disturbance storm time index (DST, or Symmetric-H (Sym-H)) \citep{Sugiura_1964}, the auroral electrojet index \citep[AE,][]{Davis_1966}, and the planetary k-index \citep[Kp,][]{Bartels_1939}.
\end{enumerate}

\begin{figure}[h]
\caption{Space weather CRIs during a geomagnetic storm on March 1, 2018. The top panel shows the impact on the electric power grid through a direct GIC measurement. The red dashed line indicates a threshold level important to power grid engineers (above which is considered a `risk.' Vertical orange lines on all plots indicate periods during which the GIC level exceeded the threshold and provide an indication of the behavior of the CRI at those important times. The variables shown are: (second panel from the top) the solar wind magnetic field z-component; (third panel from the top) the solar wind velocity; (third panel from the top) the DST/Sym-H index; and (bottom  panel) the Newell coupling function.} 
\centering
\includegraphics[width=0.55\textwidth]{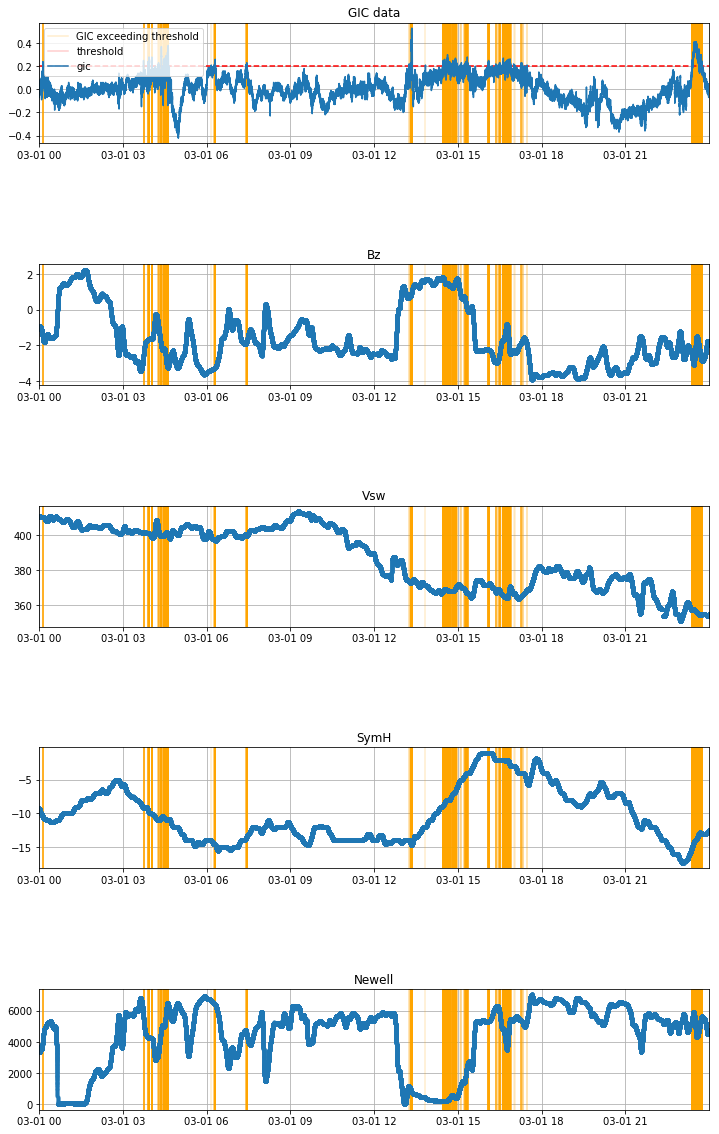}
\vspace{-10mm}
\label{space weather CRIs figure}
\end{figure}

Figure \ref{space weather CRIs figure}shows a CRI from categories 2-4 along with direct measurements of GIC (i.e., impact on the power grid). The top panel shows the GIC measurement with a red dashed line indicating a threshold level important to power grid engineers. Vertical orange lines on all plots indicate periods during which the GIC level exceeded the threshold and provide an indication of the behavior of the CRI at those important times. The variables shown are: (second panel from the top) the solar wind magnetic field z-component; (third panel from the top) the solar wind velocity; (third panel from the top) the DST/Sym-H index; and (bottom  panel) the Newell coupling function. 

Given that the currents in the Earth's atmosphere directly drive disturbances to the power grid system, the most relevant category are the proxies of the currents - the geomagnetic indices. These indices are each created by aggregating ground-based magnetometer observations. There are numerous such indices, and we will describe only the most relevant to the power grid application. The most traditional data for quantifying potential risk to the power grid by space weather is the the planetary k-index (Kp). It has long been used to communicate space weather activity to the power grid. Kp quantifies disturbances in the horizontal component of earth's magnetic field with an integer in the range 0-9 with 1 being calm and 5 or more indicating a geomagnetic storm. It is a single three-hour resolution number for the planet to proxy geomagnetic activity and many power grid models and procedures are queued to it. While Kp has proven useful, it does not provide the level of granularity needed by the power grid community because the risk is different based on region and finer location and on shorter time scales. 

Improvement is possible by using more of the information available in ground-based magnetometer measurements. This is the approach of various geomagnetic indices. The DST/Sym-H and AE indices each select a specific set of magnetometers and aggregate their data to provide a more direct indication of the atmospheric currents near the equator (Sym-H) and the auroral region (AE). These indices are provided on one-minute temporal resolution and give a more regional quantification. The Super Magnetometer Initiative (SuperMAG; \url{supermag.jhuapl.edu/} \citep{Gjerloev_2009}) provides their own versions of these indices that uses more magnetometer stations. As mentioned, power grid impacts occur on the regional level, too. Thus, a significant extension of the geomagnetic activity approach is to group magnetometer data by local time region and to create proxies that are regionally-dependent. SuperMAG provides these regional indices at one-minute resolution as well. 

The state-of-the-art would be direct observations of the power grid disruption, which are regularly collected by utilities, but seldom available for research and predictive model development. The final Space Weather CRI, therefore, are direct measurements of the induced currents on power grid transformers, GICs. Future CRI development will utilize these data to better quantify the connection between Space Weather variables and power grid risks. 


\subsection{Finance} \label{finance section}

\textit{Connection between finance and the electric power grid}:  Electricity grid and finance are tightly coupled.  Fuel costs, generation capacity costs, operating costs, transmission related costs, such as congestion pricing, investments in peak capacity, and costs related to grid infrastructure improvements and maintenance connect the two domains.

Public utility companies such as Pacific Gas Electric, Duke Energy Corp. and others are responsible for being reliable sources of electricity for individuals, private and public sectors.  Public utilities make money from investment in assets such as oil and natural gas pipelines, substations, transmission lines, etc. that are used to provide the service.  During financial crises, the finances of public utilities might be constrained due to liquidity and financing constraints, leading to decrease in investments in infrastructure, which increases the susceptibility of infrastructure.  The health and longevity of electricity grid is directly impacted by financing and the health of the economy. 

Vulnerabilities of the electric grid can also spill over to economy and depress asset values of companies, especially public utility companies.  On a macro scale, power supply interruptions directly affect the health of the economy. For large companies, the cost of a power supply interruption can escalate into the millions of dollars per hour of downtime. The U.S. cost of sustained power interruptions is \$44 billion per year in 2015, which grew by 25\% since 2002 (\cite{BerkeleyLab}). On a micro scale, power supply interruptions affect the health of companies and can precipitate their default. For example, Southern California Edison agreed to pay \$650,000 settlement for the 2011 blackout.  Due to colossal losses of \$30 billion during catastrophic wildfires caused by Pacific Gas \& Electric company (PG\&E) equipment that further led to severe power supply interruptions, PG\&E filed for Chapter 11 bankruptcy in 2019.  

In addition, energy and finance domains are clearly linked through the costs of commodities, i.e., natural gas, coal, and crude oil, which are standard inputs for electricity generation.

\textit{Existing CRIs for finance-power grid connections}: All measures are constructed using daily data.  Volatility Indicator (VIX) is a proxy for financial instability.  Public Utility indicator is an index of major US public utility companies.  These companies are traded daily on NYSE, major U.S. stock exchange.  Futures and spot contracts for crude oil, natural gals, coal, and electricity are traded daily on New York Mercantile Exchange.

\begin{enumerate}
\item \textit{Volatility Indicator (VIX)}

The CBOE Volatility Index (VIX) is a measure of expected stock market volatility based on S\&P 500 index options over the next 30 days. It is a measure of implied volatility, and specifically, model-free implied volatility. It is calculated by the Chicago Board Options Exchange (CBOE) and is often termed as the "fear index" or "fear gauge". Market participants use the VIX to measure the level of risk, fear, or stress in the market when making investment decisions.

Mathematically, the VIX is calculated as a 30-day expectation of volatility given by a weighted portfolio of out-of-the-money European options on the S\&P 500 index. The formula is as follow:
\begin{equation}
   VIX= \sqrt{\frac{2e^{r\tau}}{\tau}\left( \int_{0}^{F} \frac{P(K)}{K^2} dK+\int_{F}^{\infty} \frac{C(K)}{K^2}dK \right) } 
\end{equation}

Where $\tau$ is the number of average days in a month (30 days), $r$ is the risk-free rate, $F$ is the 30-day forward price on the S\&P 500, and  $P(K)$ and $C(K)$ are prices for puts and calls with strike $K$ and 30 days to maturity.

While the formula is theoretically complex, the intuition is as follows. It estimates the expected volatility of the S\&P 500 index by aggregating the weighted prices of multiple SPX puts and calls over a wide range of strike prices. 

 In our data sample of daily CBOE S\&P500 Volatility Index (Figure \ref{fig:vix}), VIX ranges from the lowest 9.14 on 11/3/2017 to highest 80.86 on 11/20/2008.  Note, the spike in VIX is associated with financial market turmoil, which happened during the peak of the financial crisis of 2008.  VIX also spiked during other financial crises such as the Asian Financial crisis of 1997, the Internet bubble of 2000, and the most recent COVID-19 crisis (March 2020).

\begin{figure}[h]
    \centering
    \includegraphics[width = 0.6\textwidth]{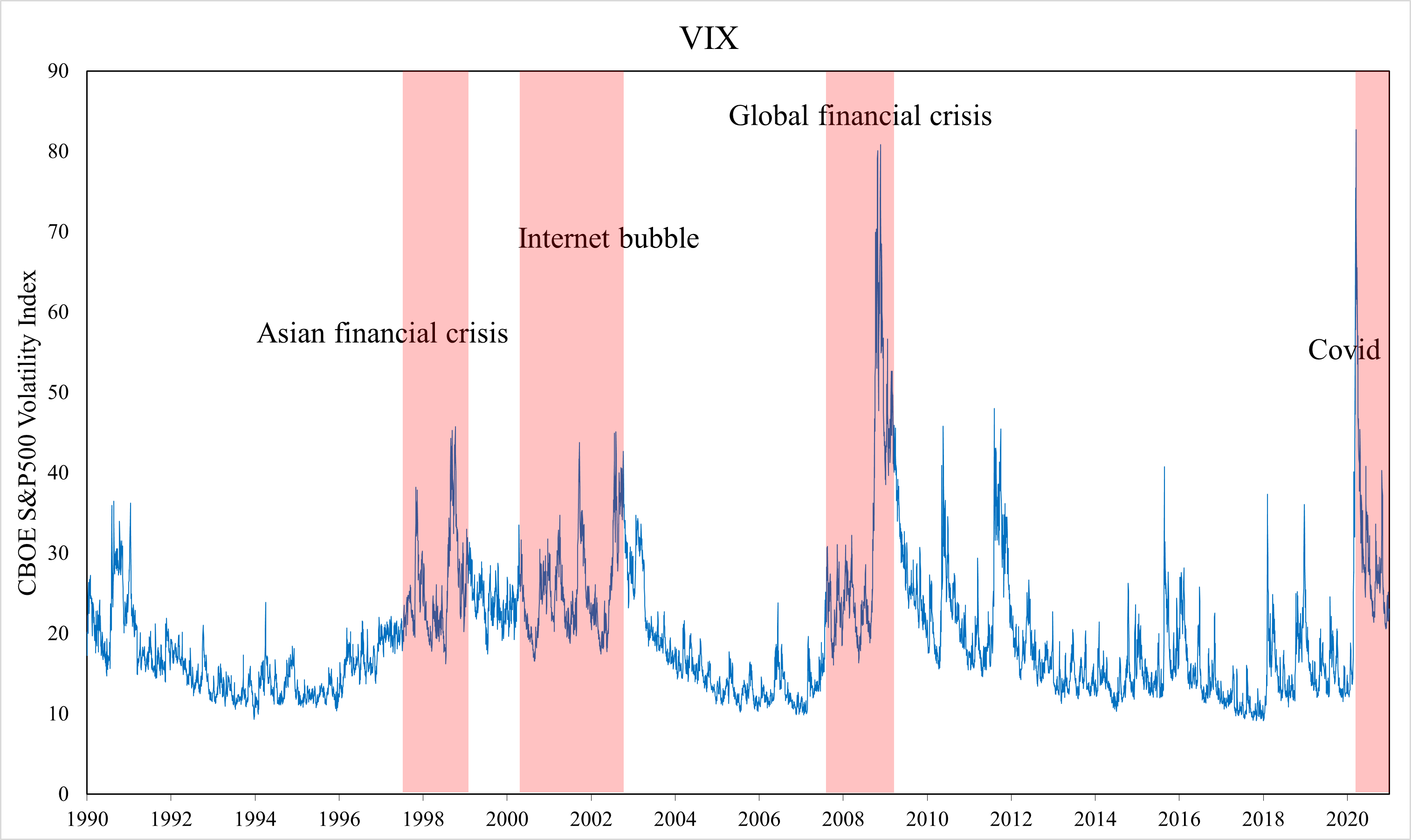}
    \caption{Daily CBOE S\&P500 Volatility Index (VIX) using closing data from 1/2/1990 to 12/30/2020.}
    \label{fig:vix}
\end{figure}

\item \textit{Public Utility Indicator}

Public utility company is an organization that maintains the infrastructure for public service. Those companies provide a set of services such as coal, electricity, natural gas, and water. 

To construct the critical risk indicator for public utility firms, we collect daily stock prices for five major public utility companies which include: Southern California Edison, Pacific Gas \& Electric, Duke Energy Corp, Consolidated Edison, and CMS Energy Corporation. We then calculate daily returns of each company using their daily closing prices and take the equal weighted average of each company's return to construct the aggregate index for public utility firms. This index serves as an indicator of public utility industry and reflects the daily stock performance of major public utility firms.

Figure \ref{fig:utility}depicts daily returns for the index of five major public utility companies from 1/2/1990 to 12/30/2020. The companies are exposed to the state of the economy and had the largest changes in value around Internet bubble and the 2008 financial crisis.  Public utility stocks are also exposed to natural disaster risk. Stock price for public utility stocks is directly impacted by natural disasters such as the wildfires in California and hurricanes on the East coast of U.S in 2018 and more recent Western U.S. wildfires in the summer of 2020. 

\begin{figure}[h]
    \centering
    \includegraphics[width = 0.6\textwidth]{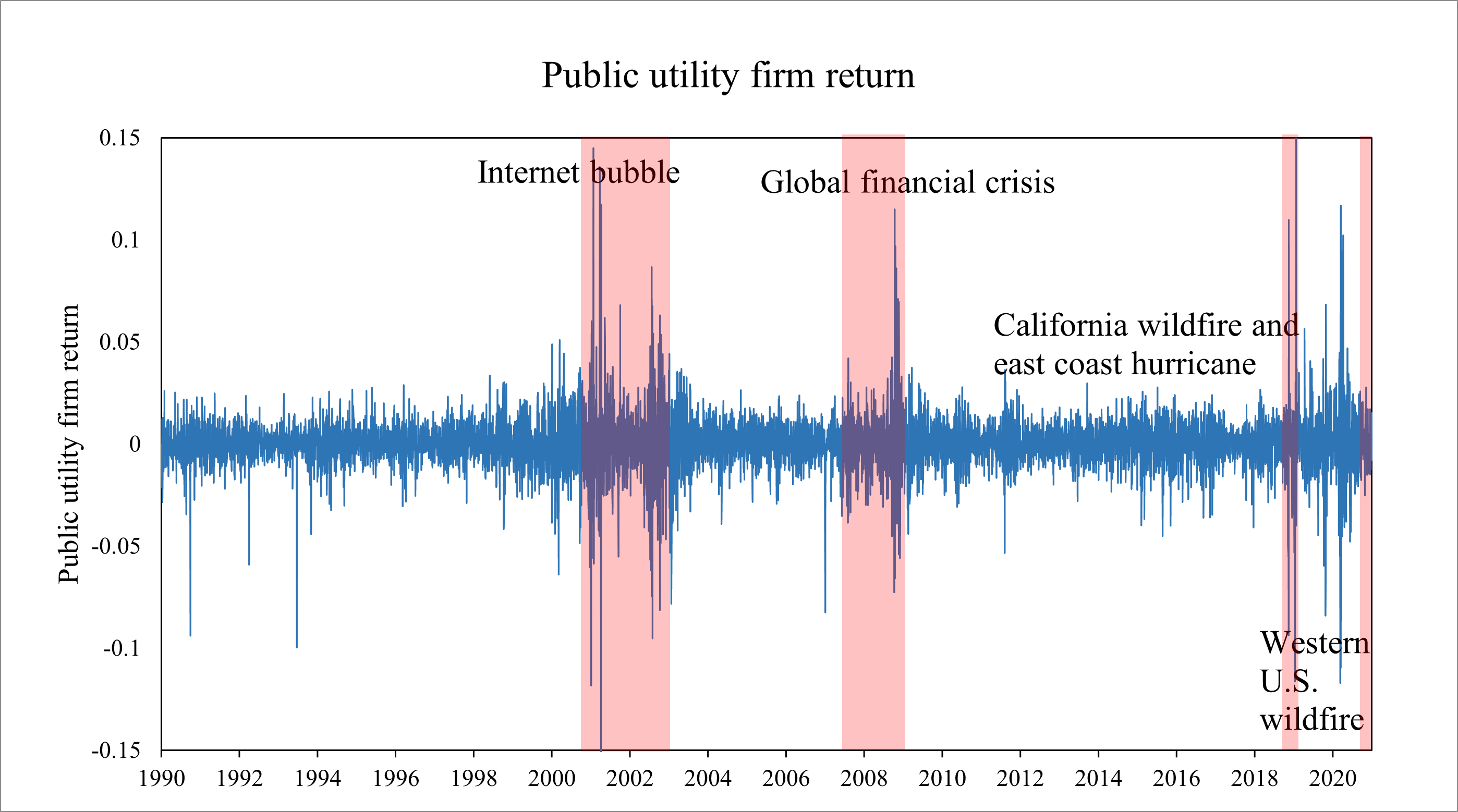}
    \caption{Daily prices for the index of five major public utility companies from 1/2/1990 to 12/30/2020.}
    \label{fig:utility}
\end{figure}

\begin{figure}[h]
    \centering
    \includegraphics[width = 0.9\textwidth]{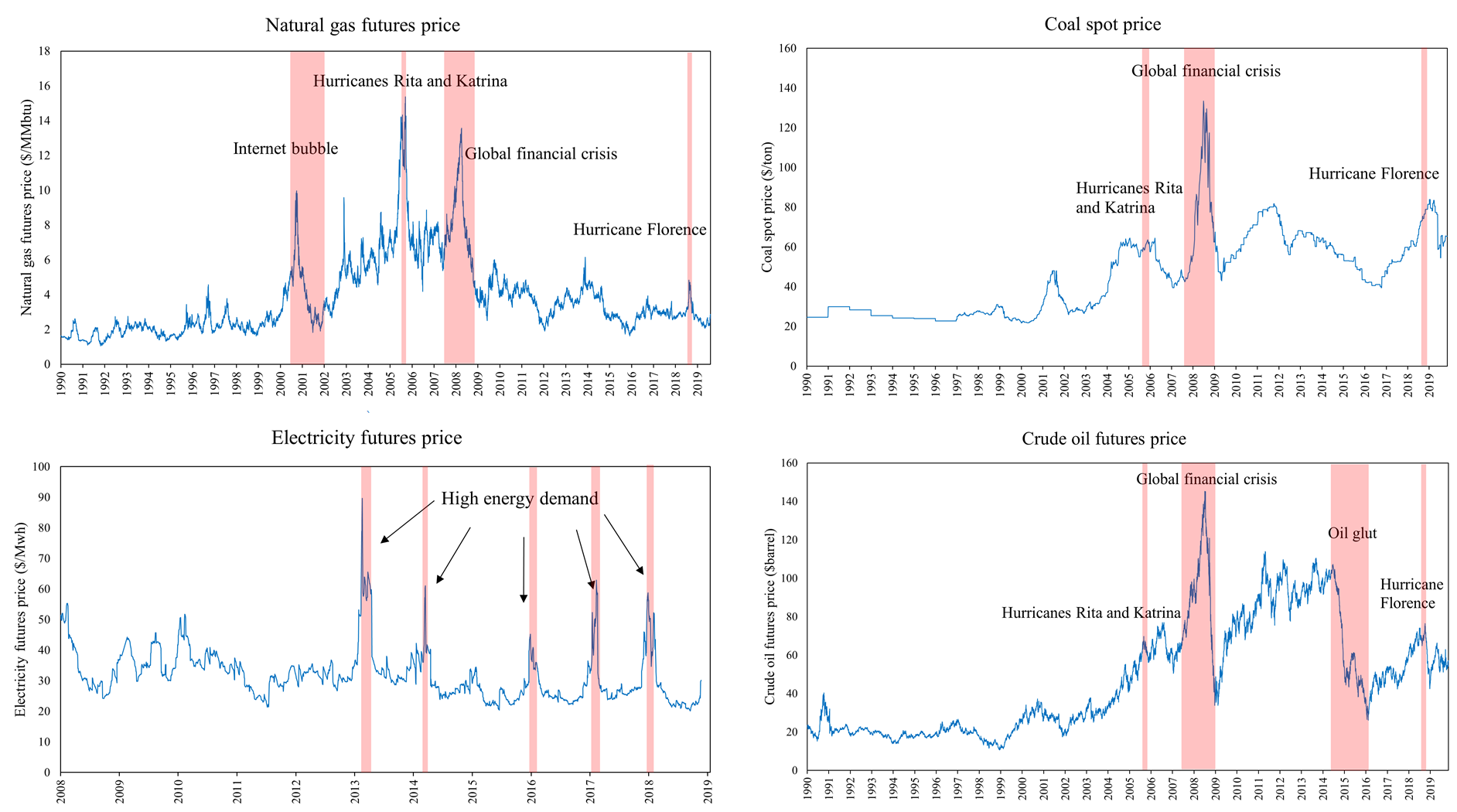}
    \caption{Daily prices for natural gas, coal, crude oil, and electricity.  Coal and crude oil data is available from 1/2/1990 to 11/5/2019.  Natural gas futures are available from 4/30/1990 to 11/5/2019 and electricity futures data is from 12/15/2008 to 11/5/2019.}
    \label{fig:commodity}
\end{figure}

\item \textit{Crude Oil Indicator}

Crude oil is a global commodity that trades in markets around the world, both as spot oil and via derivatives contracts. Crude oil is the most important and commonly traded commodity in the world as it is the primary source of energy production. To construct the indicator for crude oil, we use the futures price of crude oil as an index since Central banks and the International Monetary Fund (IMF) mainly use oil futures contract prices as their gauge for the level of oil prices. Specifically, we use the daily price of CME Crude Oil Future as the indicator. 

As demand for oil goes up, crude oil futures increase in price.  The largest run-up of crude oil prices was right before the global financial crisis in 2008 followed by the largest decline in our time period (from \$140 per barrell to \$40 per barrell).  In 2014-2015 the world experience the oil glut where a serious surplus of crude oil resulted in the plunge of oil prices during this time period. Crude oil prices are also related to natural disasters and spiked during Hurricanes Katrina (2005), Rita (2005), and Florence (2018) (see Figure \ref{fig:commodity}).  

\item \textit{Natural Gas Indicator}

Natural Gas is a traded commodity with many industrial and commercial applications. In the United States it is traded as a futures contract on the New York Mercantile Exchange. The price of natural gas is mainly driven by supply and demand fundamentals. It may also be linked to the price of crude oil and petroleum products. To construct the indicator for natural gas, we use the Henry Hub Natural Gas Futures price as an index.

As demand for natural gas goes up, natural gas futures increase in price.  The largest run-up of natural gas prices was right before the global financial crisis in 2008 followed by the largest decline in our time period.  In addition to financial crises (Internet bubble of 2000 and global financial crisis of 2008\add{), }natural gas prices are impacted by natural disasters such as hurricanes Katrina (2005), Rita (2005), and Florence (2018)(see Figure \ref{fig:commodity}). 

\item \textit{Coal Indicator}

To construct the indicator for coal, we use the Thermal Coal Historical Spot Price as an index. Spot price is the price for a one-time open market transaction for immediate delivery purchased on the spot at current market rates. Coal prices have historically been lower and more stable than oil and gas prices. 

Demand for coal has resulted in strong price movements in the commodity itself. Before the 2008 Global Financial Crisis, prices for coal experienced a major uptrend, going from \$50 per short ton in 2006 to almost \$140 per short ton in 2008. Coal prices are also impacted by natural disasters such as hurricanes Katrina (2005), Rita (2005), and Florence (2018)(see Figure \ref{fig:commodity}). 

\item \textit{Electricity Indicator}

Electricity is a commodity capable of being bought, sold, and traded. Electricity futures and other derivatives can help market participants manage, or hedge, price risks in a competitive electricity market. Futures contracts are legally binding that call for the future delivery of the commodity. To construct the indicator for electricity, we use the PJM Western Hub Real-Time Off-Peak Calendar-Month 5 MW Futures price as an index. 

Electricity prices are a function of conditions of the economy, demand for electricity, and prices of electricity inputs such as natural gas, crude oil, and coal.  During our sample period we show that electricity prices spiked during the financial crisis (2008) and during high energy demand caused by cold weather in the beginning of 2013, 2014, 2016, 2017, and 2018 (see Figure \ref{fig:commodity}).  2014 and 2017 saw the spike in natural gas prices.  2013, 2014, and 2018 saw the spike in crude oil futures prices.

While electricity price is listed as a finance CRI, it also is an electric energy CRI. Not only do CRIs generate risks that can spill over into other domains but also many CRIs don't conveniently fit in siloed domains. Through this network analysis approach, the role of CRIs across multiple domains becomes increasingly apparent.

\end{enumerate}

\subsection{Summary of CRIs} \label{summary of cris section}
In this section we summarize top CRIs from each domain (climate, hydrology, agriculture, ecology, space weather, and finance) that relate to electric power grid. For each domain we provide Jupyter Notebooks to illustrate and provide a foundation for further exploration of the domain-specific CRIs outlined in this manuscript (see the electronic supplementary material). These are useful tools to facilitate interaction between data scientists and domain scientists.  

\begin{center}
\begin{tabular}{ c c c c }
Domain & CRI & Affected by Grid & Affects Grid \\ 
\hline
 Climate & Anomalies (rainfall, temperature) & no & yes \\
 Climate & Standard Precipitation Index (SPI) & no & yes \\
 Climate & Anomalies of number of days a criteria is met (e.g., $> 1$mm; $\leq 0$\degree C) & no & yes \\
 Hydrology & Streamflow & yes & yes \\
 Hydrology & Drought indices & no & yes \\
 Hydrology & Groundwater levels & yes & yes \\
 Agriculture & Irrigation demand & no & yes \\
 Agriculture & Crop biomass production & yes & yes \\
 Agriculture & Vegetation Index (EVI) & yes & yes \\
 Ecology & Population abundance (Living Planet Index) & yes & yes \\
 Ecology & Bird abundance (USGS Breeding Bird Survey) & yes & yes \\
 Ecology & Biodiversity (Shannon and Simpson indices) & yes & yes \\
 Space Weather & Kp Index & no & yes \\
 Space Weather & Global SuperMAG indices (SMR and SME) & no & yes \\
 Space Weather & Regional SuperMAG indices (SMR and SME) & no & yes \\
 Space Weather & Power Grid Geomagnetically Induced Currents (GICs) & no & yes \\
 Finance & Volatility Indicator (VIX) & yes & yes \\
 Finance & Public Utility Indicator & yes & yes \\
 Finance & Crude Oil Indicator & yes & yes \\
 Finance & Natural Gas Indicator & yes & yes \\
 Finance & Coal Indicator & yes & yes \\
 Finance & Electricity Indicator & yes & yes \\
 
\end{tabular}
\label{summary cris table}
\end{center}

\section{Electric Energy} \label{energy section}
\par
\textit{Realization of risk in the power grid}: The \change{culmination of}{negative outcome associated with} risk in the electrical grid \change{results in}{is} a \textit{power supply interruption}. The\remove{se} risk\remove{s} could originate wholly from within the electric energy domain or as a result of spillovers from other domains (as discussed in section \ref{Critical Risk Indicators (CRIs) by Domain}).  A \textit{power supply interruption} can be defined as the total loss of electric power on one or more normally energized conductors to one or more customers connected to the distribution portion of the system \citep{Subcommittee2012}. This does not include any of the power quality issues such as: sags, swells, impulses, or harmonics \citep{Subcommittee2012}. 
In contrast, an outage is the loss of ability of a component to deliver power, which may or may not result in an interruption \citep{Subcommittee2012}. The severity of power supply interruptions also depends on the domain where the risk is emanating from, for example risks from the space weather domain as a result of magnetic storms tend to result in major electric grid disturbance while risks emanating from ecology - as a result of interference by animals, e.g., squirrels and birds, are comparatively less severe.   

\par
\textit{Existing Electric Energy CRIs}: CRIs in the electric energy domain relate to either the \add{potential for or} severity of a power supply interruption\remove{ or the risks of a potential power supply interruption}. Examples of existing electric energy CRIs include:
\begin{enumerate}
    \item \textit{System Average Interruption Duration Index (SAIDI)}: One of the most widely used metric for quantifying disturbances on the power grid is the SAIDI. According to the \textit{IEEE Guide for Electric Power Distribution Reliability Indices} \citep{Subcommittee2012}, this metric is used to quantify the amount of time, on average, customers' electricity is disrupted for longer than five minutes. SAIDI is defined as:
\begin{equation}
\label{eq_SAIDI}
    \textrm{SAIDI} = \frac{\sum_{}^{}r_{i}N_{i}}{N}
\end{equation}
    \item \textit{System Average Interruption Frequency Index (SAIFI)}: SAIFI on the other hand indicates how often the average customer experiences a sustained interruption over a predefined period of time. it is defined as: 
\begin{equation}
\label{eq_SAIFI}
    \textrm{SAIFI} = \frac{\sum_{}^{}N_{i}}{N}
\end{equation}
where $r_{i}$ is the duration of each interruption $i$, $N_{i}$ is the amount of customers affected, and $N$ is the total number of customers being serviced   \citep{Subcommittee2012}. 
    \item \textit{Reserve Margin}: Reserve margin is an energy system metric used to quantify the adequacy of generation resources to satisfy demand \citep{NERC2013}. As opposed to aforementioned CRIs, the reserve margin can indicate a potential to result in a power supply interruption. It measures (as shown in Eq.(\ref{eq_RM})) the percentage of available generation that exceeds the peak demand where "available generation" is the maximum supply available. 
\begin{equation}\label{eq_RM}
   RM  = \frac{G_{av} - D}{D} 
\end{equation}
where $RM$ is the reserve margin, $G_{av}$ is the available generation capacity and $D$ is the peak demand. 
\end{enumerate}
The System Average Interruption Duration Index (SAIDI) and the System Average Interruption Frequency Index (SAIFI) are examples of indices that relate to a failure of the power system since they can only be calculated after a failure has occurred. Both SAIDI and SAIFI are very useful for measuring the resulting risk outcome on the electrical grid and they can be very useful in developing strategies for improving grid reliability. 
\par
Reserve margin serves various purposes depending on the context within which it is used. In the US Electricity sector, planning reserve margin, PRM, is used to quantify the capacity build-out required to meet an adequacy threshold. North American Electric Reliability Corporation (NERC) uses it to evaluate resource planning decisions uniquely for each region in the US. In the absence of regional targets, NERC assigns a 15\% reserve margin for thermal systems and 10\% for hydro systems \citep{Nerc2020}. An issue with the reserve margin is that the metric does not quantify the specific impact of each generator in meeting demand. The definition and evaluation of the reserve margin is changing to adapt to the transformation of the electrical grid by accounting for penetration of more variable resources, such as wind and solar. It can, therefore, be modified to account for resource availability and forced outage rates. The reserve margin can also point to supply security\remove{ risk}, especially in cases where generation imports are necessary to satisfy peak demand. 

\begin{figure}[htp]
\caption{Electric Energy CRIs for ISO New England between 2016 and 2018. The variables shown are: SAIDI and the reserve margin (top panel); SAIFI and the reserve margin (bottom panel). The red lines indicate the severity of the power supply interruption (SAIDI or SAIFI). The blue curves indicate the reserve margin based on historical data of generation and demand for each day within the time period \citep{CommonwealthofMassachusetts2020}.}
\centering
\includegraphics[width=17.5cm]{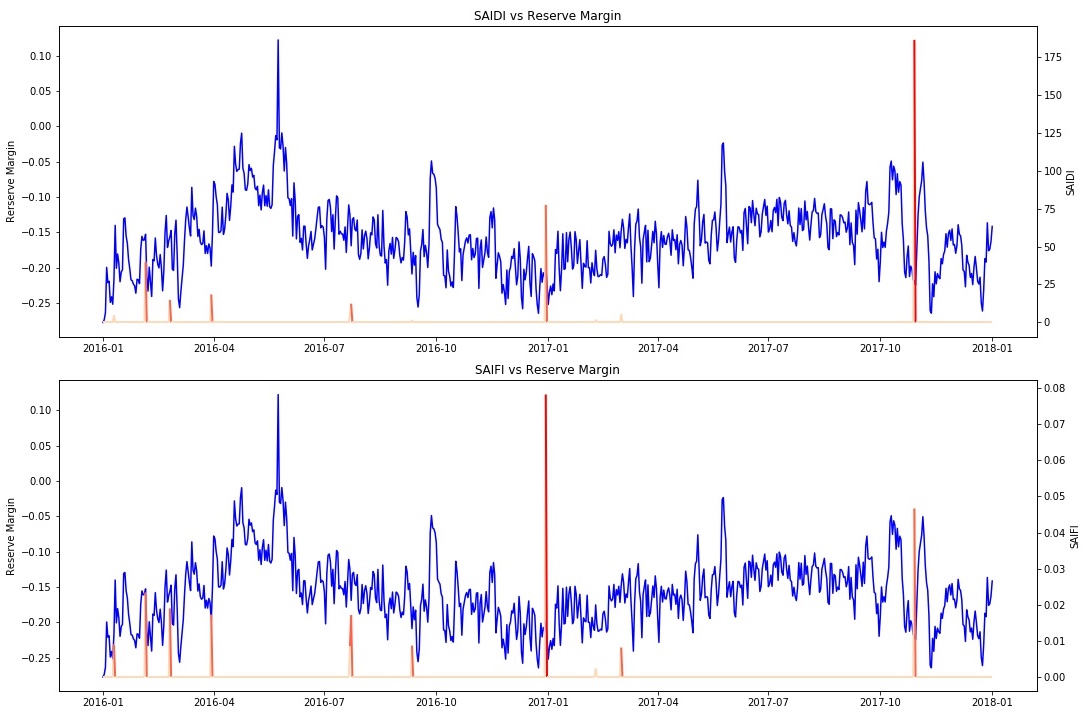}
\label{Energy CRIs figure}
\end{figure}

Figure \ref{Energy CRIs figure}shows the plot of SAIDI and SAIFI against the reserve margin\footnote{Here, the reserve margin has been calculated using the actual generation and demand (so the generation does not necessarily represent the \textit{maximum available capacity.})}. Negative values of the reserve margin indicates that external generation is needed to satisfy demand. A decrease in reserve margin is observed around January, 2017 and this coincides with a spike in SAIFI and SAIDI. 

SAIDI, SAIFI, and Reserve Margin are provided as illustrative examples of electric energy CRIs. Please note that numerous other risk and reliability indices exist, such as the expected energy not supplied (EENS) and its associated costs, i.e., customer interruption costs.
\par
\textit{Available data sets for calculating existing Electric Energy CRIs}: Reliability data consisting of SAIDI and SAIFI are reported annually (available from 2013 - 2018) by the US Energy Information Administration \citep{U.S.EnergyInformationAdministration2019}. However, outage data reported in the OE-417 Electric Disturbance report (for major outages available 2000 till present) \citep{OfficeofCybersecurity}  and those collected at the state/utility level (an example is the Massachusetts Outage Data \citep{CommonwealthofMassachusetts2020} usually provide higher spatial and temporal resolution for these indices. 

\section{Systemic Risk Indicators (SRI)} \label{systemic risk}

This survey shows that the electrical grid and its resilience are not defined and operated in a closed system.  Critical risk indicators in the electric energy domain are directly impacted by other domains such as climate, ecology, hydrology, agriculture, space weather, and finance.  We identified CRIs in each of the domains that are directly related to electrical grid vulnerability. Although an existing CRI in one domain may be important for modeling risk to the power grid, individual CRIs may only be important during specified scenarios or time frames (e.g., space\change{-}{}weather CRIs may only exhibit strong signals during solar storms). 

Moreover, in addition to considering bilateral relationships with each human-natural domain and the electric energy domain, it is important to look at systemic interconnections between CRIs among domains. We borrow the concept of systemic risk and systemic risk measures from finance literature.  In finance, systemic risk measures the risk of financial system instability, which is caused or exacerbated by idiosyncratic events or catastrophic conditions in financial intermediaries \citep{Billio2012}. It is the risk that the collapse of one financial institution could cause other connected financial institutions to fail and harm the real economy as a whole.  In our setting, systemic risk measures will capture the health of the interconnected human-natural system and the interdependencies between CRIs in each of the domain; we refer to the trans-domain systemic risk measures as Systemic Risk Indicators (SRIs). Below we provide narrative for additional human-natural domain connections and develop a framework for assessing systemic risk and building systemic risk measures for human-natural domains.  

\subsection{Additional Examples of Domain Interconnections} 
Climate risk indicators directly relate with other domains such as ecology, hydrology, and agriculture. Excess temperatures above a certain threshold are known to favor the growth of certain crops. Rainfall is important to farmers, especially those who practice rain-fed agriculture, and so monthly anomalies of rainfall are a relevant CRI for agriculture. Monthly (or seasonal) aggregation of rainfall may, however, not be the best CRI for rain-fed agriculture as the sequence of the rainfall events during the month (season) matter as much as, if not more than, the total rainfall in that period. With the same amount of rain during the crop growing season, outcomes to the crop health may be drastically different if the rainfall is evenly spread in different days, rather than if it rains in one or two extreme events. Therefore the number of wet days (days above a certain precipitation threshold, e.g., 1mm) may be a better CRI to agricultural yields.

Much of the same reasoning can be applied to ecology, as the activity and developmental rates of wildlife and their habitats vary according to daily and seasonal weather \citep[temperature, precipitation, and wind speeds;][]{doostan_statistical_2019} or accumulated heat \citep[growing degree days;][]{murray2020}. Therefore, weather data can account for biases in species abundance data due to variation in detection probability. Threshold-based analyses may be good indicators for identifying key increases or reductions of species abundance that result in changes to biodiversity that can be accompanied with high risk outcomes in other domains. Broader scale spatial and temporal climate anomalies may be more applicable to measures of biodiversity change. Regional changes in climate can change habitat suitability. This can lead to some species declining in abundance and others increasing depending on their adaptability to the changes in climate. Overall this causes changes in species compositions and shifts in species distributions \citep{doi:10.1111/1365-2435.13095}.

Additionally, hydrologic risk can spill over to other domains such as aquatic ecosystem health \citep{Falke2011}. The drought-induced soil moisture deficit affects vegetation productivity and crop yield. Irrigation demand under droughts is generally met with groundwater pumping, which requires electricity \citep{Scott_2013}. As indicated in the Hydrology section, the CRI is "drought", a longer-term building up process that starts with sustained deficit in rainfall (meteorological drought) that may turn into agricultural drought, hydrological drought and below-average hydropower production. Climate CRIs relevant to hydrology are indicators such as the Standardized Precipitation Index (SPI) that indicates the build-up of rainfall deficit (or excess) for the past 3 to 24 months. While climate inputs are the driving forces, the hydrologic system strongly modulates the input signal. The variable that is more directly related to stakeholders, such as reservoir operators, and civil infrastructure is streamflow. From the perspective of monitoring and forecasting future risks, monthly outlook of streamflow distributions can be a valuable CRI.

Despite their positive impacts on agricultural productivity, electrification can also generate negative spillovers. The availability of electricity is often accompanied by environmental costs, for example, groundwater over-exploitation \citep{badiani2018electricity}. Policies that are related to electricity price (e.g., subsidy) are therefore particularly important, for example, in India, where groundwater irrigates 70\% of irrigated agricultural land \citep{badiani2018electricity}. The process of electricity generation (e.g., burning coal, oil or natural gas, hydropower) is also not without economic, social, and environmental impacts. For example, hydropower operation is determined by scheduling that typically includes several river basins \citep{gonzalezmultipurpose_res}. The scheduling alters water levels and downstream water flow patterns \citep{castelletti2008water}, affecting habitats of different flora and fauna. Although part of the electricity generated by hydropower is also used for irrigation pumping machines, there may exist a competition of water use between power generation with irrigation itself \citep[e.g., when streamflow is limited;][]{gonzalezmultipurpose_res}.

Electric energy domain can also affect agriculture through at least two different mechanisms. First, the U.S. electric power sector was responsible for roughly 30\% of total U.S. energy-related CO$_2$ emissions \citep{EIA_energy_review}. These emissions may create negative feedback to agricultural production due to its effect on climate while offering CO$_2$ fertilizer effects. Second, if power supply interruption occurs, the industrial-sector (including agriculture) may experience significant reduction in productivity as its outage-per-customer cost (USD 3,253 per one-hour outage) is significantly higher than commercial (USD 886) and residential sectors \citep[USD 2.7; based on currency value in 2002;][]{lacommare2004understanding}.

In summary, there are many explicit interconnections among human-natural domains that have been studied.  Each domain has a unique risk profile and specific CRIs.  We next move to developing a framework for connecting these domains and assessing resulting systemic risk.

\begin{figure}[h]
\centering
\subfigure[Domain CRIs Monthly Time Series, Jan 2006 to Dec 2018. ]{\label{fig:a}\includegraphics[width=0.49\textwidth]{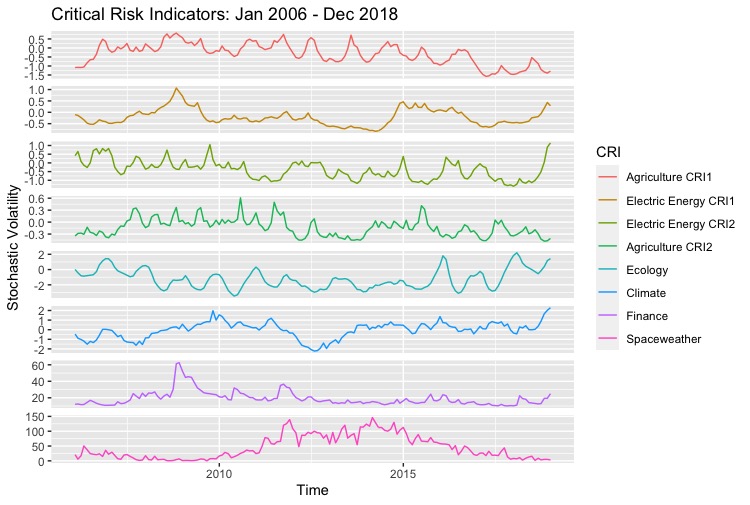}}
\subfigure[SRI Granger-Causality Network from VAR(1) Model. Directed arrows indicates temporal dependency (past $\rightarrow$ present). Arrow width indicates magnitude of the relationship; positive vs.\ negative relationships\ are colored blue vs.\ red. ]{\label{fig:b}\includegraphics[width=0.49\textwidth]{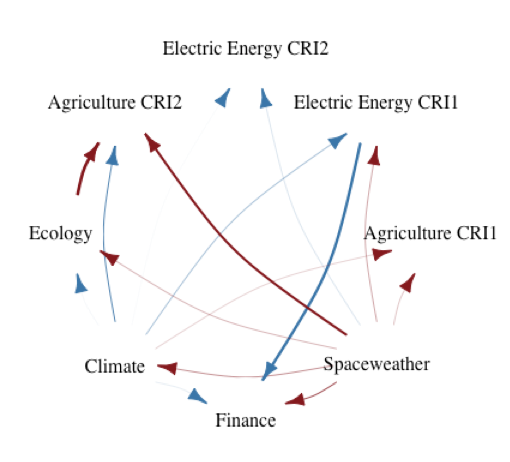}}
\caption{Initial SRI Analysis Results}
\label{fig:sri_var}
\end{figure}

\subsection{Systemic Risk Indicators for Human-Natural Domain Interconnections}

We broadly define trans-domain or holistic systemic risk as any set of circumstances that threatens the stability of our society and natural world; in our case, systemic risk concerns the overall stability and reliability of our power grid system at large. Systemic risk arises endogenously from a nexus of interconnected processes or systems. It is a function of the connections between the fundamental structure of these diverse components, and thus causality and dynamics play a central role. We propose using a dynamic model of CRIs from many domains (Figure \ref{fig:sri_var}a), in order to (1) define a trans-domain risk nexus represented as a network or graph, and (2) construct SRIs as functions of connectivity from that network.

As a starting point, we introduce the vector autoregression (VAR) model for dynamically modeling CRIs. It remains the standard-bearer for macroeconomic forecasting \citep{sims1980} and is widely applied in numerous fields, including climatology, neuroscience, and signal processing. 
Suppose we have computed a historical sequence of $i = 1,\ldots, d$ distinct CRIs, within and across several domains, and that they are aligned with common time index $t = 1,\ldots, n$. For now, also suppose the CRIs are approximately stationary over this time period and mean-centered.
Let $\textbf{y}_t$ denote the $d$ dimensional vectorization of these CRIs at time $t$, such that the $i$th component of $\textbf{y}_t$ corresponds to $\textrm{CRI}_i(t)$.
In a VAR model, the series $\textbf{y}_t$ is modeled as a function of its own past value $\textbf{y}_{t-1}$, which may be standardized for better parameter estimation. More precisely,
\vspace{-1mm}
\begin{equation}
    \textbf{y}_t  =  \Phi \textbf{y}_{t-1}  +  \textbf{a}_{t}
\end{equation}
where $\Phi$ is the ${d \times d}$ autoregressive parameter matrix, 
$\textbf{a}_t$ denotes a $d$-dimensional mean-zero white noise (serially uncorrelated) vector time series with $d\times d$ nonsingular (contemporaneous) covariance matrix ${\Sigma}_a$. 
This VAR model also allows easy extensions for incorporating additional lagged values (e.g., $\textbf{y}_{t-2}$, etc.), straightforward forecasting at multiple horizons including forecast intervals, simple forecast updating, and (impulse) response function analysis \citep{tsay2013multivariate}. Further extensions are possible to account for non-stationarity and multi-level spatial-temporal resolution. To account for potential false discoveries across a large collection of CRIs, we can utilize any number of multiple testing methods \citep{benjamini1995controlling} that control for False Discovery Rate (FDR). Based on our estimated parameters, we can construct a network of inter-temporal dependencies (Granger causality) based on significance thresholds and utilize the network summary statistics, such as connectivity, as a SRI. This graph can either be unweighted where an edge between two nodes indicate existence of a significant inter-temporal dependency, or can be weighted by the magnitude of the dependency. In finance, for example, Granger causality provide measurements of directional connection between financial institutions over two consecutive time periods. Using Granger causality network, \citet{Billio2012} found that during the global financial crisis of 2007-2009, returns of banks and insurers seem to have more signiﬁcant impact on the returns of hedge funds and broker/dealers than vice versa. 

After proper estimation of inter-temporal relationships (e.g., Granger causality), we propose the following network-based SRIs based on Granger-causality network of interconnected CRIs: eigenvector centrality and degrees of Granger causality.  The \textit{eigenvector centrality\/} measures the importance of a CRI in a network by assigning relative scores to CRIs based on how connected they are to the rest of the network. First, define the adjacency matrix $A$ as the matrix with elements: $\left[A\right]_{ji} \ \ =\ \ \left( j\rightarrow i\right)\ $, where an edge is determined by if a statistically significant Granger-causal relationship exists between two nodes. This edge can either be weighted or unweighted.
%
The eigenvector centrality measure is the eigenvector $v$ of the
adjacency matrix associated with eigenvalue 1, i.e., in matrix form: $Av=v\ .$
\textit{Degree of Granger causality\/} (DGC) is the fraction of statistically significant Granger-causality relationships among all $N(N\!-\!1)$ pairs of $N$ CRIs, and is one way to summarize connectivity over the entire graph. If the adjacency matrix has non-negative entries, a unique solution is guaranteed to exist by the Perron-Frobenius theorem.

As an illustrative example, consider the set of $d = 8$ CRIs across six human-natural domains measured monthly from January 2006 to December 2018.  The domains (agriculture, electric energy, ecology, climate, finance, and space-weather) and weekly time-series of individual CRIs are shown in (Figure \ref{fig:sri_var}a). Note that some domains can have more than one CRI. In this case, we use two CRI measures each for agriculture and electric energy domains. From the estimate of $\Phi$, the inter-connectivity can be determined by statistically significant off-diagonal coefficients, which produce a model-estimated trans-domain nexus (Figure \ref{fig:sri_var}b). Specifically, we construct a Granger-causality network from VAR(1) model, where directed arrows indicate temporal dependence, arrow width indicates magnitude of the relationship, and color (blue vs. red) indicates positive vs. negative relationships. We observe that space weather is an exogenous CRI that affects all domains but is not impacted by any other domain. In comparison, the finance domain is strongly affected by other domains but bears no impact on others (no out-going arrows). Some domains, such as electric energy, agriculture, ecology, and climate are more inter-dependent, and a shock to one CRI can easily propagate to others. The total number of significant Granger-causal relationships is 15, giving a DGC of ~26.8\% which indicates moderate connectivity. While Figure \ref{fig:cri_dag} shows a general schematic and nexus of interconnections, Figure \ref{fig:sri_var}b provides a specific example of how a Granger-causality network for SRIs can be constructed and interpreted.

An alternative measure to consider is \textit{cosine similarity} \citep{Girardi2018} between CRIs. As an example from financial institutions, \citet{Girardi2018} found that similar institutions' asset holdings (those with high cosine similarity) lead to massive joint sales that leads to subsequent drop in asset prices, during and after large disasters such as Hurricanes Harvey and Rita. Other systemic risk measures that focus on probabilities of loss and network methods have been introduced in the finance and economics literature. \cite{bisias2012survey} provides a survey of the following systemic risk measures: marginal expected shortfall (MES), SRISK, turbulence measure, network connectedness method (PCA, and Granger Causality), and volatility measure. It is important to emphasize that no one measure will be best for capturing systemic risk in human natural systems, as the performance of individual metrics can vary at different spatial and temporal scales and depending on the specific components measured \citep{Bakkensen2017982}. Frameworks for combining multiple metrics have been proposed to comprehensively assess the resilience of energy systems \citep[e.g.,][]{Roege2014249, Gatto2020}. Similarly, a dashboard of different systemic risk measures can provide a more holistic understanding of trans-domain interactions.

\section{Acknowledgements} 
Funding was provided by the NSF Harnessing the Data Revolution (HDR) program, "Collaborative Research: Predictive Risk Investigation SysteM (PRISM) for Multi-layer Dynamic Interconnection Analysis" (awards \#1940160, 2023755, 1940176, 1940190, 1940208, 1940223, 1940276, 1940291, and 1940696). R. McGranaghan was partially supported under the NSF Convergence Accelerator Award to the Convergence Hub for the Exploration of Space Science (CHESS) team (NSF Award Number: 1937152). We would like to thank Suoan Gao (UMASS Amherst) for research assistance.


\bibliographystyle{spbasic}

\section{Bibliography}
\renewcommand{\bibsection}{}
\bibliography{PRISM_CRIs_Survey}

\section{Electronic supplementary material}

Jupyter Notebooks that illustrate the domain-specific CRIs outlined in our manuscript, including sample data and exploratory code, can be found at \url{https://github.com/rmcgranaghan/Critical-Risk-Indicators-CRIs-for-the-electric-power-grid/}.

\section{Ethics declarations}
{\bf Conflict of interest}
The authors declare that they have no conflicts of interest.



%

\end{document}